\documentclass[aps,pre,
amsmath,
amssymb,
showpacs,
floatfix,
reprint,
linenumbers,
superscriptaddress,
]{revtex4-1}
\usepackage{wasysym}
\usepackage{graphicx}
\usepackage{ulem}

\usepackage{CJK}
\usepackage{xcolor}

\renewcommand{\eqref}[1]{~(\ref{#1})}

\newcommand{\upd}{\mathrm{d}}

\begin{document}

%\setpagewiselinenumbers
%\modulolinenumbers[3]
%\linenumbers
\begin{CJK*}{GB}{gbsn} % Use default fonts from CJK (see below)

\title{Arbitrary-order Hilbert spectral analysis for  time series possessing
scaling statistics: a comparison study with detrended fluctuation analysis and wavelet leaders}

\author{Y.X. Huang (»ÆÓÀÏé)}
\email{yongxianghuang@gmail.com}
\affiliation{Shanghai Institute of Applied
         Mathematics and Mechanics,
        Shanghai University,  Shanghai 200072,  China}
        
\affiliation{Shanghai Key Laboratory of Mechanics in Energy and Environment Engineering, Yanchang Road, Shanghai 200072, China
}
\affiliation{E-Institutes of Shanghai Universities, Shanghai University, Shanghai 200072, China
}

\affiliation{Universit\'e Lille Nord de France, F-59044 Lille, France}
\affiliation{USTL, LOG, F-62930 Wimereux, France}
\affiliation{CNRS, UMR 8187, F-62930 Wimereux, France}
\affiliation{Environmental Hydroacoustics Laboratory, Universit\'e Libre de
Bruxelles, Avenue F-D. Roosevelt 50 - CP 194/5, B-1050 Brussels, Belgium}%Lines
%break automatically
%or can be forced with \\

\author{F. G. Schmitt}%
 \email{francois.schmitt@univ-lille1.fr}
\affiliation{Univ Lille Nord de France}
\affiliation{USTL, LOG, F-62930 Wimereux, France}
\affiliation{CNRS, UMR 8187, F-62930 Wimereux, France}

\author{J.-P. Hermand}
\affiliation{Environmental Hydroacoustics Laboratory, Universit\'e Libre de
Bruxelles, Avenue F-D. Roosevelt 50 - CP 194/5, B-1050 Brussels, Belgium}%Lines

\author{Y. Gagne}
\affiliation{
LEGI, CNRS/UJF/INPG, UMR 5519, F-38041 Grenoble, France}

 \author{Z.M. Lu (¬־Ã÷)}
\affiliation{Shanghai Institute of Applied
         Mathematics and Mechanics,
        Shanghai University,  Shanghai 200072,  China}
        
\affiliation{Shanghai Key Laboratory of Mechanics in Energy and Environment Engineering, Yanchang Road, Shanghai 200072, China
}
\affiliation{E-Institutes of Shanghai Universities, Shanghai University, Shanghai 200072, China
}
%\email{zmlu@shu.edu.cn}
\author{Y.L. Liu (ÁõÓî½)}
%\email{ylliu@staff.shu.edu.cn}
\affiliation{Shanghai Institute of Applied
         Mathematics and Mechanics,
        Shanghai University,  Shanghai 200072,  China}
        
\affiliation{Shanghai Key Laboratory of Mechanics in Energy and Environment Engineering, Yanchang Road, Shanghai 200072, China
}
\date{\today}% It is always \today, today,

\begin{abstract}
In this paper we present an extended version of Hilbert-Huang transform, namely arbitrary-order Hilbert spectral analysis, to characterize the scale-invariant properties of a time series directly in an 
amplitude-frequency space. We first show numerically that due to a nonlinear distortion, traditional methods require high-order harmonic components to represent nonlinear processes, except for the 
Hilbert-based method. This will lead to an artificial energy flux from the low-frequency (large scale) to the high-frequency (small scale) part. Thus the power law, if it exists, is contaminated. We then 
compare the Hilbert method with structure functions (SF), detrended fluctuation analysis (DFA), and wavelet leader (WL) by analyzing fractional Brownian motion and synthesized multifractal time series. 
For the former simulation, we find that all methods provide comparable results. For the latter simulation, we perform simulations with an intermittent parameter $ \mu= 0.15$. We find that the SF 
underestimates scaling exponent when $q>3$. The Hilbert method provides a slight underestimation when $q>5$. However, both DFA and WL overestimate the scaling exponents when $q>5$. It 
seems that Hilbert and DFA methods provide better singularity spectra than SF and WL. We finally apply all methods to a passive scalar (temperature) data obtained from a jet experiment with a Taylor¡¯s 
microscale Reynolds number $Re_lambda \simeq 250$. Due to the presence of strong ramp-cliff structures, the SF fails to detect the power law behavior. For the traditional method, the ramp-cliff structure causes a 
serious artificial energy flux from the low-frequency (large scale) to the high-frequency (small scale) part. Thus DFA and WL underestimate the scaling exponents. However, the Hilbert method provides 
scaling exponents $\xi_\theta(q)$ quite close to the one for longitudinal velocity, indicating a less intermittent passive scalar field than what was believed before.
\end{abstract}

%{Time series analysis}

\maketitle

\end{CJK*}

% main text
\section{Introduction}
Multifractal properties have been found in many fields, such as
turbulence~\cite{Anselmet1984,Frisch1995,Lohse2010},
rainfall~\cite{Schertzer1987,Schmitt1998,Delima1999,venugopal2006}, financial time
series~\cite{Ghashghaie1996,Schmitt1999,Lux2001,Calvet2002},
physiology~\cite{Ivanov1999}, etc. 
Conventionally,  multifractal properties of such time series are
characterized by the  scaling exponents $\zeta(q)$, which are extracted by structure
function (SF) analysis:
%\begin{equation}
$\Delta V_{\ell}^q\sim \ell^{\zeta(q)}$,
%\end{equation}
where $\Delta V_{\ell}=\vert V(x+\ell)-V(x) \vert$ are the increment for
 scale separation $\ell$, and $\zeta(q)$ is a nonlinear function~\cite{Anselmet1984,Frisch1995,Monin1971}.
 The function  $\zeta(q)$ is  linear for monoscaling processes and nonlinear
 for multifractal processes.
  We may also mention the detrended fluctuation analysis {(DFA)}~\cite{Peng1994PRE,Hu2001,Bashan2008}
or the multifractal detrended fluctuation analysis~\cite{Kantelhardt2002a}, which are  sometimes
also employed for scaling time series analysis.
 The DFA method is similar to SFs since it involves increments
and  characterizes the scale invariance in the physical domain.

Other widely used methods are  wavelet-based methods, e.g.  wavelet
transform modulus maxima (WTMM),    wavelet leader (WL), or gradient modulus wavelet projection (GMWP), to extract the
scaling exponents from a scaling time series \cite{Muzy1991PRL,Muzy1993PRE,
Arrault1997PRL,Rodrigues2001,Farge1996IEEE,Farge1992,Ghez1989JSP,
Jaffard2005wavelet,Pont2006PRE,Turiel2006JCP,Lashermes2008EPJB,Serrano2009,Lashermes2005wavelet}.
However,  as we will show in this paper, the wavelets share  the same drawback
with Fourier transform, which requires high-order harmonic components to represent 
nonlinear processes.

Some of  us have proposed recently a new methodology, namely arbitrary-order Hilbert
spectral analysis (HSA), to characterize the scale invariant properties directly
in amplitude-frequency space~\cite{Huang2008EPL,Huang2008TSI,Huang2009PHD}.
It is an extended version of the Hilbert-Huang transform (HHT), which provides a
 joint probability density function (pdf) in
an amplitude-frequency space~\cite{Huang2008EPL,Huang2009PHD}.
We have applied part of this new methodology to several different time series to
show its efficiency and validity: 
 turbulence
experimental database~\cite{Huang2008EPL}, synthesized fractional Brownian motion (fBm) time
series~\cite{Huang2008TSI}, surf zone marine turbulence
data~\cite{Schmitt2009JMS}, and river flow discharge
data~\cite{Huang2009Hydrol}. In this paper we consider in length and precisely this new
method and its validation and calibration. We first introduce this new
methodology in detail.
 We then validate and calibrate it by analyzing
a synthesized  multifractal  time series. We finally
consider a passive scalar (temperature) data set with strong ramp-cliff structures. Due to the
presence of ramp-cliff structures, the classical SF analysis
fails to detect the power law behavior \cite{Huang2010PRE}.  Additionally,  for traditional methods, such as
Fourier transform, wavelet transform, high-order harmonics is required to represent these structures and  leads to an artificial energy flux from
the large-scale part (low frequency) to the small-scale part (high frequency) \cite{Huang2010PRE}.

This paper is organized as follows. We present the definition of the
arbitrary-order Hilbert spectral
analysis in Sec.~\ref{sec:methodology}. The classical structure function
analysis, multifractal detrended fluctuation analysis and wavelet leader are also
presented in this section. We then consider a nonlinear effect by using the
classical
Duffing equation to show the artificial high-order harmonic components required
by the classical methods, for example, Fourier transform,  and wavelet transform, in Sec.
\ref{sec:nonlinear}. In Sec.~\ref{sec:data} we perform a comparison study of
the arbitrary-order HSA with other methods by analyzing a fBm simulation and a synthesized
multifractal time series.  We then present in Sec.  \ref{sec:ps} an analysis of real
temperature data obtained from a jet experiment. We finally draw
the main conclusions
in Sec.~\ref{sec:conclusion}.

\section{Methodology}\label{sec:methodology}
\subsection{Arbitrary-order Hilbert spectral  analysis}\label{sec:HHT}
\subsubsection{Empirical mode decomposition}

The most innovative part of the Hilbert-Huang transform  is the so-called
empirical mode decomposition (EMD). In the real world most of the signals are  multi-components, 
which means that different scales can  coexist
simultaneously~\cite{Cohen1995,Huang1998EMD,Huang1999EMD}. This may be considered 
as   fast oscillations superposed to slower ones  at  a  local
level~\cite{Rilling2003EMD,Flandrin2004EMDa}. Meanwhile, for decomposition  methods, a characteristic scale (CS)
is always defined implicitly or explicitly before the decomposition. For example, the CS of the classical 
Fourier analysis is a period of sine wave. The CS of wavelet transform is the shape of the mother wavelet  \cite{Huang2009PHD}.
 In the present method, the CS  is defined as the distance
between  two successive maxima (respectively minima) points. Then the so-called
intrinsic mode functions (IMF)
 are proposed to represent each  mono-component signal. An IMF  satisfies the
following two conditions:  (\romannumeral1) the difference between
the number of local extrema and the number of zero-crossings must be zero or
one; (\romannumeral2) the running mean value of the envelope
defined by the local maxima and the envelope defined by the local minima is
zero~\cite{Huang1998EMD,Huang1999EMD}.

A subpart of the  EMD algorithm, called ``sifting process,''
 is then designed to decompose  a given signal
 into several IMF modes~\cite{Huang1998EMD,Huang1999EMD,Rilling2003EMD}. The
first step of the sifting process is to identify all the local  maxima (respectively,
minima) points for a given time series  $x(t)$. Once all the local extrema 
points
are identified,    the upper  envelope $e_{\max}(t)$ and the lower envelope
$e_{\min}(t)$ are constructed, respectively, for the local  maxima and minima
points  by using  a cubic spline algorithm.   The mean between these two
envelopes is defined as
 \begin{equation}
 m_{1}(t)=\frac{e_{\max}(t)+e_{\min}(t)}{2}
 \end{equation}
 Thus the first component is  estimated by
\begin{equation}
h_1(t)=x(t)-m_{1}(t)
\end{equation}
 Ideally,  $h_1(t)$ should
be an IMF as expected.  However,  $h_1(t)$  may  not satisfy
 the above-mentioned conditions to be an IMF. The function $h_1(t)$ is then
taken as a new time series and  this sifting process is repeated $j$ times, until
$h_{1j}(t)$ is an IMF
\begin{equation}
h_{1j}(t)=h_{1(j-1)}(t)-m_{1j}(t)
\end{equation}
The first IMF component $C_{1}(t)$ is then written as
\begin{equation}
C_{1}(t)=h_{1j}(t)
\end{equation}
and the residual $r_1(t)$ as
\begin{equation}
r_{1}(t)=x(t)-C_{1}(t)
\end{equation}
from the data $x(t)$. The sifting procedure is then repeated
on the residual  until  $r_n(t)$ becomes a monotonic function or at most has one
local extreme point, which means   that no more IMF can be extracted  from
$r_n(t)$.  There are finally  $n-1$
IMF modes with one residual $r_n(t)$. The original signal $x(t)$ is   rewritten 
at the end of the process as
\begin{equation}
x(t)=\sum_{i=1}^{n-1}C_i(t)+r_{n}(t)
\end{equation}
To guarantee
 that the IMF modes retain enough physical sense, a certain stopping criterion
has to be introduced to stop
the sifting process properly.
 Different types of stopping criteria have been introduced by
 several
authors~\cite{Huang1998EMD,Huang1999EMD,Rilling2003EMD,Huang2003b,Huang2005EMDa}.
 The first stopping
 criterion is a  Cauchy-type convergence criterion. We introduce  the standard
deviation (SD),  defined for two successive sifting processes as
\begin{equation}
\mathrm{SD}=\frac{\sum_{t=0}^{T}\vert
h_{i(j-1)}(t)-h_{j}(t)\vert^2}{\sum_{t=0}^{T} h_{i(j-1)}^2(t)}
\end{equation}
If a calculated SD is smaller than a given value, then the sifting stops,
and gives an IMF.
A typical value  is $0.2\sim 0.3$, proposed based on Huang et al.\rq{}s experiences \cite{Huang1998EMD,Huang1999EMD}.
Another widely used criterion is based on three thresholds $\alpha$, $\theta_1$, and
$\theta_2$, which are designed  to guarantee globally small fluctuations
  meanwhile taking into account locally large
excursions~\cite{Rilling2003EMD}.
The mode amplitude and evaluation function are  given as
\begin{subequations}
\begin{equation}
a(t)=\frac{e_{\max}(t)-e_{\min}(t)}{2}
\end{equation}
and
\begin{equation}
\sigma(t)=\vert m(t)/a(t)\vert
\end{equation}
\end{subequations}
So that the sifting is iterated until $\sigma(t)<\theta_1$ for some prescribed
fraction $1-\alpha$ of the total duration, while $\sigma(t)<\theta_2$ for
the remaining fraction. The typical values proposed by~\citet{Rilling2003EMD}  are
$\alpha \approx 0.05$, $\theta_1
\approx 0.05$ and $\theta_2 \approx 10 \,\theta_1$, respectively based on
their experience.
In our practice, if one of these criteria is satisfied,  then
the sifting process will stop. We also set a maximal iteration number (e.g., $300$) to avoid over-decomposing the time series.

The above-described EMD algorithm performs the decomposition on a very local
level
in the physical domain without \textit{a priori} basis. This means that the
present decomposition is   \textit{a posteriori}: The basis is induced by the
data itself~\cite{Huang1998EMD,Huang1999EMD,Flandrin2004EMDa}. It is thus a scale-based decomposition.
Since its introduction, this method has attracted large interests in various
research fields: {waves}~\cite{Schmitt2009JMS,Hwang2003,Veltcheva2004},
biological applications~\cite{Echeverria2001,Balocchi2004,Ponomarenko2005},
financial studies~\cite{Huang2003a}, meteorology and climate
 studies~\cite{Huang2009Hydrol,Coughlin2004,Janosi2005,Molla2006a,Sole2007,
Wu2007}, mechanical engineering~\cite{Loh2001,Chen2004},
acoustics~\cite{Loutridis2005}, aquatic environment~\cite{Schmitt2007}, and
turbulence~\cite{Huang2008EPL}, to quote a few.
 More detail about the EMD algorithm can be found in
 several methodological papers~\cite{Huang1998EMD,Huang1999EMD,
 Rilling2003EMD,Flandrin2004EMDa,Flandrin2004EMDb,Huang2005EMDa}.

 \subsubsection{Hilbert spectral analysis}

 After having extracted the IMF modes, one can apply the associated Hilbert
spectral analysis to each  component $C_i$  in order to extract the
energy time-frequency information
 from the data~\cite{Huang1998EMD,Huang1999EMD,Long1995}.
 The Hilbert transform of a function ${C}(t)$ is written as
\begin{equation}
\tilde {C}(t)=\frac{1}{\pi}{P}\int{\frac{C(t')}{t-t'} \upd
t'}\label{eq:Hilbert}
\end{equation}
where $P$ means the Cauchy principle
value~\cite{Cohen1995,Huang1998EMD,Long1995,Flandrin1998}.  For each mode function
${C}_{i}(t)$, one can  then  construct the analytical
signal~\cite{Cohen1995,Flandrin1998}, $\mathbb{C}_i(t)$, as
\begin{equation}
\mathbb{C}_i(t)=C_i(t)+j
\tilde{C}_i(t)=\mathcal{A}_i(t)\mathrm{e}^{j\theta_i(t)}\label{eq:analytic}
\end{equation}
where
\begin{equation}
\left\{ \begin{array}{lll}
&\mathcal{A}_i(t)=[C_i(t)^2+\tilde{C}_i^2(t)]^{1/2}\\
& \theta_i(t)=\arctan
\left(\frac{\tilde{C}_i(t)}{C_i(t)}\right)
\end{array}\right.
\end{equation}
Hence the  instantaneous frequency is defined as
\begin{equation}
\omega_{i}=\frac{1}{2\pi} \frac{\upd\theta_{i}(t)}{\upd t}
\end{equation}
The original signal is finally represented [excluding the residual $r_n(t)$] as
\begin{equation}
  x(t)=\mathrm{R}\sum_{i=1}^{N}{\mathcal{A}_{i}(t)e^{j\theta_{i}(t)}}=
  \mathrm{R}\sum_{i=1}^{N}{\mathcal{A}_{i}(t)e^{j\int{\omega_{i}(t)}d t}}
\end{equation}
where "R" means real part. The above procedure is the classical Hilbert spectral analysis \cite{Cohen1995,Flandrin1998}. The combination of EMD and HSA is thus called Hilbert-Huang transform by some authors \cite{Huang2005EMDa}. The
Hilbert-Huang transform can
be taken as a generalization of the Fourier transform,  since it allows
frequency modulation
and amplitude modulation simultaneously. The
Hilbert spectrum, $H(\omega,t)=\mathcal{A}^2(\omega,t)$, is designed to
represent the energy in a time-frequency representation~\cite{Long1995}. We
further
can define the Hilbert marginal spectrum as
\begin{equation}
  h(\omega)=\int_{0}^{+\infty}{H(\omega,t)\upd t}\label{eq:marginal1}
\end{equation}
This is similar with the Fourier spectrum, and can be interpreted as the energy
associated with each  frequency.  However, we underline the fact that
the  definition  of frequency here is
different from the definition in the Fourier
frame.
Thus the interpretation of the Hilbert marginal spectrum should be  given  more caution \cite{Huang1998EMD,Huang1999EMD}.  

%%%%%
\subsubsection{Arbitrary-order Hilbert spectral analysis}\label{sec:HSA}
We  can also define the joint pdf $p(\omega,\mathcal{A})$ of the instantaneous
frequency $\omega$ and the amplitude $\mathcal{A}$ for
{each} of these IMF modes
\cite{Huang2008EPL,Huang2008TSI,Long1995,Huang2009PHD}. The Hilbert
marginal spectrum Eq.\eqref{eq:marginal1}  is then rewritten as
 \begin{equation}
 h(\omega)=\int_0^{+\infty}
 p(\omega,\mathcal{A})\mathcal{A}^2 \upd \mathcal{A} \label{eq:marginal2}
\end{equation}
The  above definition is no more than the second-order statistical moment. This
constatation
has led some of us to recently generalize this approach to arbitrary-order moment
$q\ge0$~\cite{Huang2008EPL,Huang2008TSI,Huang2009PHD}
 \begin{equation}
 \mathcal{L}_q(\omega)=\int_0^{+\infty}
 p(\omega,\mathcal{A})\mathcal{A}^q \upd \mathcal{A}
  \label{eq:arbitrary}
\end{equation}
 In case of scale invariance, we have
  \begin{equation}
   \mathcal{L}_q(\omega)  \sim \omega^{-\xi(q)}
  \end{equation}
  in which $\xi(q)$ is the Hilbert-based scaling exponent function.  Due to
the integration operator,  $\xi(q)-1$ can be associated
with $\zeta(q)$ from SF analysis \cite{Huang2008EPL,Huang2009PHD}.

 A limitation of the Hilbert-based
method we proposed here  is that it   lacks  the ability  to consider $q<0$ \footnote{In fact, Eq.\eqref{eq:arbitrary}   converges when $q\ge -1$. However, in practice, we only consider the case $q\ge0$. }. In
other words,  similarly  with the
SF analysis, it has
no resolution on the right part of the singularity spectrum. The main drawback
of the Hilbert-based method is its absence of solid
  theoretical ground, since the EMD part is almost empirical~\cite{Huang2005EMDa}.
It has been
  found experimentally that the method,  especially for the HSA, is
statistically stable
  with different stopping criteria~\cite{Huang2003b}.  Recently,
  Flandrin {et al.}  have obtained new theoretical results on
  the EMD method
\cite{Flandrin2004EMDa,Flandrin2004EMDb,Rilling2006,Rilling2008,Rilling2009}.
  However, more theoretical work is still needed to fully mathematically
understand
  this method.

\subsection{Structure function analysis}\label{sec:SF}

The conventional way to extract scaling exponents is the classical SF
 analysis, which has been proposed in the field of turbulence and
is now quite classical for intermittency studies~\cite{Monin1971}. The $q${th}
order SF is written as

\begin{equation}
  S_q(\ell)=\langle \vert \Delta x_{\ell}(t) \vert^q \rangle \sim
\ell^{\zeta(q)}
\end{equation}
where  $\Delta x_{\ell}(t)=x(t+\ell)-x(t)$ and $\ell$ is the time separation. The scaling
exponent $\zeta(q)$ characterizes
 the fluctuation statistic at all scales; it is linear for monofractal processes
 such as fractional Brownian motion, and nonlinear and concave (as a second
Laplace characteristic function) for multifractal
processes~\cite{Schertzer1997}.
This approach
 has been widely used in turbulent
research \cite{Anselmet1984,Frisch1995,Monin1971} and also other research fields
 \cite{Schmittbuhl1995JGR,Schmitt1995GRL,Schmitt1999}. However, the increment operation acts a filter and thus SF characterizes 
 the scale-invariant properties in an indirect way; see  detailed  discussion in Refs. \cite{Huang2010PRE,Huang2009PHD}.

As we have shown
elsewhere,  the increment operation in SF acts a filter and is a global
operation. It thus measures the scale invariant property in an indirect way. It is also found that it is strongly influenced by energetic
large scale
structures \cite{Huang2010PRE,Huang2009PHD}. Therefore the SF
analysis is not suitable for those data which possess energetic large  scale
structures.  We will show an example of passive scalar turbulence data with
strong ramp-cliff structures in Sec. \ref{sec:ps}. More discussion can be found in Refs. \cite{Huang2010PRE,Huang2009PHD}.

%%%%DFA
\subsection{Multifractal detrended fluctuation analysis}
DFA was first introduced by \citet{Peng1994PRE}
to study the scaling properties of DNA sequence, in which only the second-order
moment $q=2$  was  considered. Later this was generalized into a multifractal version
by considering the arbitrary order $q$, namely multifractal detrended
fluctuation
analysis (MFDFA) \cite{Kantelhardt2002a,Oswicecimka2006PRE}. It  then  became a 
more  common technique  for scaling data analysis
\cite{Peng1994PRE,Heneghan2000PRE,Hu2001,Kantelhardt2002a,Chen2002PRE,
KoscielnyBunde2006,Sadegh2006JSM,Oswicecimka2006PRE,Bardet2008,Zhang2008a,Bashan2008}.
For a given discrete time series $x(i)$, $i=1\cdots N$, we first estimate its
cumulative function
\begin{equation}
Y(j)=\sum_{i=1}^{j} \left(x(i)- \overline{x} \right),\quad j=1,\cdots N
\end{equation}
where $\overline{x}$ is the mean value of $x$. We then  divide it into
$M_n$ segments of length $n$ ($n<N$) starting from both the beginning and
the end of the time series. Each segment $v$ has its own local trend that
can be approximated by fitting a $p$th-order polynomial $P_{v}^{p}$  which
is
removed from the data. The variances for all the segments  $v$
and for all segment lengths $n$ are then calculated by

\begin{equation}
F^2(v,n)=\frac{1}{n}\sum_{j=1}^{n} \{ Y[(v-1)n+j]-P_{v}^{p}(j) \}^2
\end{equation}
The $q$th-order fluctuation function is then defined as

\begin{equation}
{F^q}(n)=\left(  \frac{1}{2M_n} \sum_{v=1}^{2M_n} \left[ F^2(v,n) \right]^{q/2}
\right)^{1/q}
\end{equation}
For discussion convenience, we redefine the $q$th-order
fluctuation function as

\begin{equation}
\mathcal{F}_q(n)={F^q}(n)^q\label{eq:DFA}
\end{equation}
In case of scale invariance, we have power law scaling within a significant
range of $n$

\begin{equation}
{\mathcal{F}_q}(n)  \sim n^{h(q)}
\end{equation}
in which $h(q)$ is the corresponding scaling exponent  function.

%%%wavelet
\subsection{Discrete wavelet transform and wavelet leaders}

 Wavelets have  been  widely used in data analysis and turbulence research
\cite{Muzy1991PRL,Mallat1992singularity,Farge1992,Muzy1993PRE,
Farge1996IEEE,Arrault1997PRL,Mallat1999wavelet,Rodrigues2001,
Jaffard2005wavelet,Wendt2007,
Lashermes2008EPJB,Serrano2009}. 
Several wavelet-based methods  have been  proposed by several researchers to extract the scaling exponents from a scaling time series, for example,
wavelet coefficients (WC), WTMM \cite{Muzy1991PRL,Mallat1992singularity,Muzy1993PRE},
WL \cite{Jaffard2005wavelet,Wendt2007,Lashermes2008EPJB}, etc. We   consider here WC
 and WL. 

 The discrete wavelet transform (DWT)
is defined as
\begin{equation}
\psi(k,j)=\int_{\mathbb{R}}x(t)\varphi\left(  2^{-j}t-k  \right)\upd t
\end{equation}
where $\varphi$ is the chosen wavelet, $\psi(k,j)$ is the wavelet coefficient,
$k$ is the position index, $j$ is the scale index, and $\ell=2^j$ is the
corresponding
scale \cite{Daubechies1992,Mallat1999wavelet}.
The first way to detect the scale-invariant properties is   to consider 
the wavelet coefficients
\begin{equation}
Z_q(j)=\langle \vert \psi(k,j) \vert^q \rangle\sim 2^{j\tau(q)}
\end{equation}
where $\tau(q)$  are  the corresponding scaling exponents.

 Every discrete wavelet coefficient
 $\psi(k,j)$ can be associated with the dyadic interval $\varrho(k,j)$
 \begin{equation}
 \varrho(k,j)=[2^jk,2^j(k+1))
 \end{equation}
 Thus the wavelet  coefficients  can be represented as $\psi(\varrho)=\psi(k,j)$.
Wavelet leaders are  defined  as

\begin{equation}
l(k,j)=\sup_{\varrho'\subset 3 \varrho(k,j),j'\le j} \vert \psi(\varrho')\vert
\end{equation}
 where $3 \varrho(k,j)=\varrho(k-1,j)\cup \varrho(k,j)\cup \varrho(k+1,j)$
\cite{Jaffard2005wavelet,Lashermes2005wavelet,Wendt2007}. Thus power law
behavior is expected

\begin{equation}
\mathbb{Z}_q(j)= \langle l(k,j)^q \rangle\sim 2^{j\tau(q)}
\end{equation}
in which $\tau(q)$ is the corresponding scaling exponent.
   Its efficiency has been shown for various types  of  data set 
\cite{Jaffard2005wavelet,Wendt2007,
 Lashermes2008EPJB,Serrano2009,Lashermes2005wavelet}.

 Let us recall some  previous  comparison studies between WTMM, MFDFA and WL. 
\citet{Oswicecimka2006PRE} performed a comparison study between WTMM and MFDFA by analyzing synthesized data. They stated that
the MFDFA provides a better estimation of singularity spectrum than WTMM. \citet{Jaffard2005wavelet} stated that WL provides a 
better singularity spectrum than WTMM.  \citet{Serrano2009} performed a comparison study between MFDFA and WL. They found 
that WL performs better than MFDFA. However, for a short time series, MFDFA is proposed to extract multifractal spectrum.
 A detailed  comparison 
 can be found in Ref. \cite{Jaffard2005wavelet},  \cite{Oswicecimka2006PRE}, \cite{Serrano2009}, respectively, for WTMM and WL, MFDFA and WTMM, and WL and MFDFA.

However, we argue here that DWT violates two facts of the time-frequency representation of a time series. First, the scale of a 
time series from complex system, for example, turbulent flows, is continuous in  a  statistical sense, but not discrete on several scales 
\cite{Huang2008EPL,Huang2009PHD}.  The other one is that for a certain scale, it  may   not exist all the time 
\cite{Flandrin1998,Huang1998EMD,Huang2009PHD}; see also the discussion in the next section. Thus to represent a signal by using a DWT is not 
consistent with the physical aspect.

%%% nonlinear effect
\section{Nonlinear effects}\label{sec:nonlinear}

\begin{figure}[!htb]
%\centering
\includegraphics[width=0.95\linewidth]{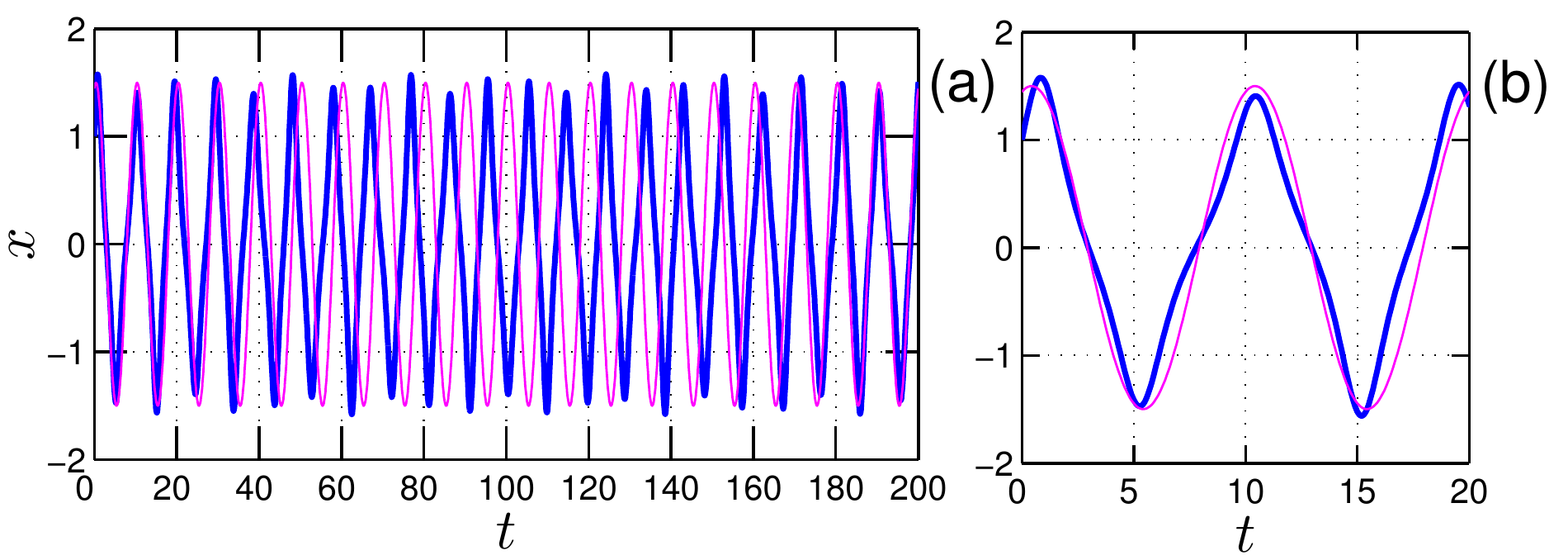}
\caption{(Color online) (a) a fifth order numerical solution (thick solid line) for Duffing
equation, 
(b)  An enlarged  portion. For comparison, a sine wave with the same mean
frequency is
also shown as a  thin solid line. The departure from a pure sine wave profile
is the result of nonlinear interactions,  which are  nonlinear distortion.
}\label{fig:Duffing}
\end{figure}

\begin{figure}[!htb]
%\centering

\includegraphics[width=0.85\linewidth]{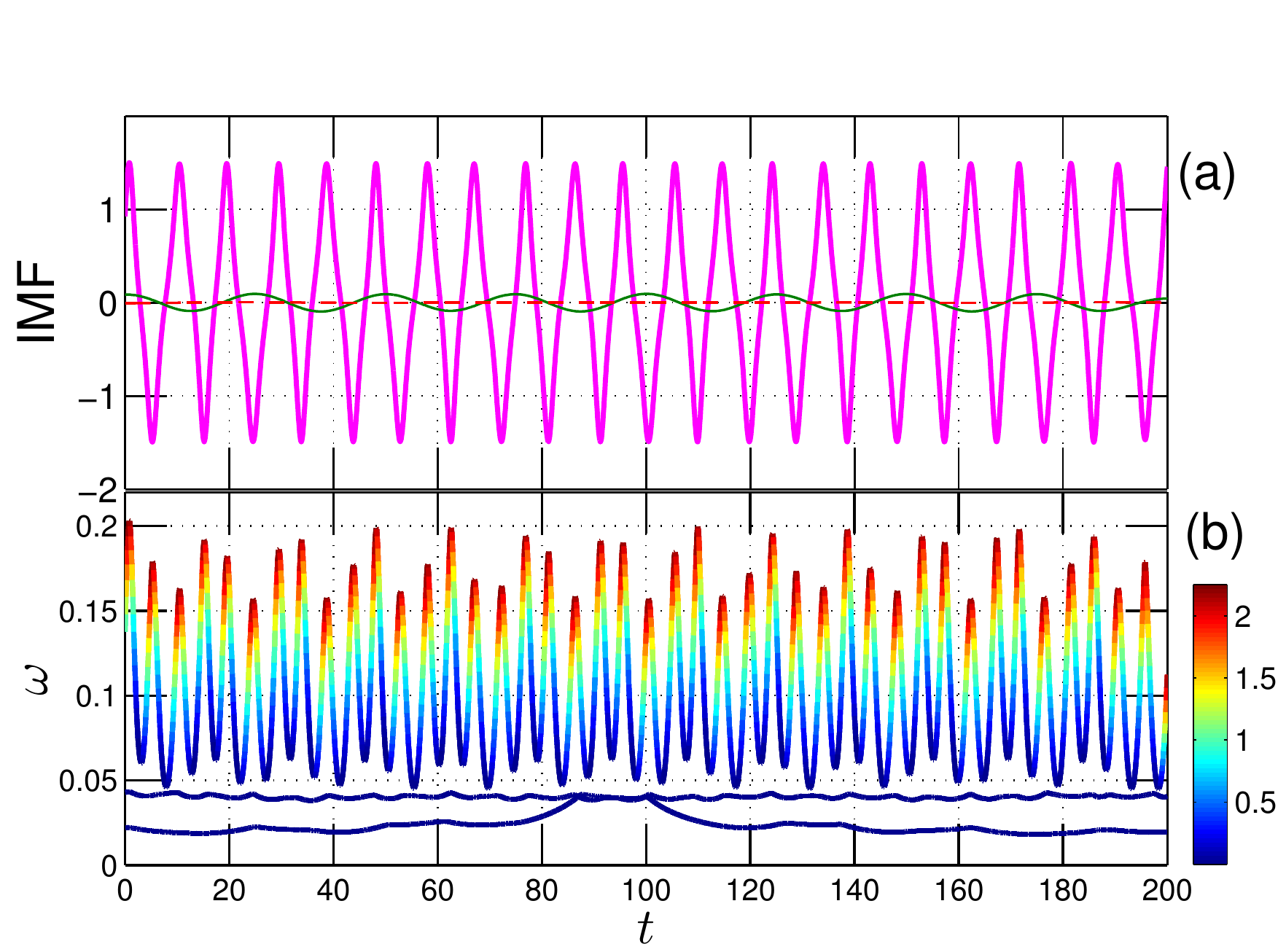}
\caption{(Color online) (a) The first three IMF modes from EMD; (b) the corresponding 
instantaneous
 frequency $\omega$ from Hilbert spectral
analysis. Note  the variation of the instantaneous frequency within
one period. It is  an  intrawave-frequency-modulation, which corresponds to  a 
nonlinear interaction. The instantaneous energy  is encoded as  a 
color.}\label{fig:IF}
\end{figure}

\begin{figure}[!htb]
\centering
\includegraphics[width=0.85\linewidth]{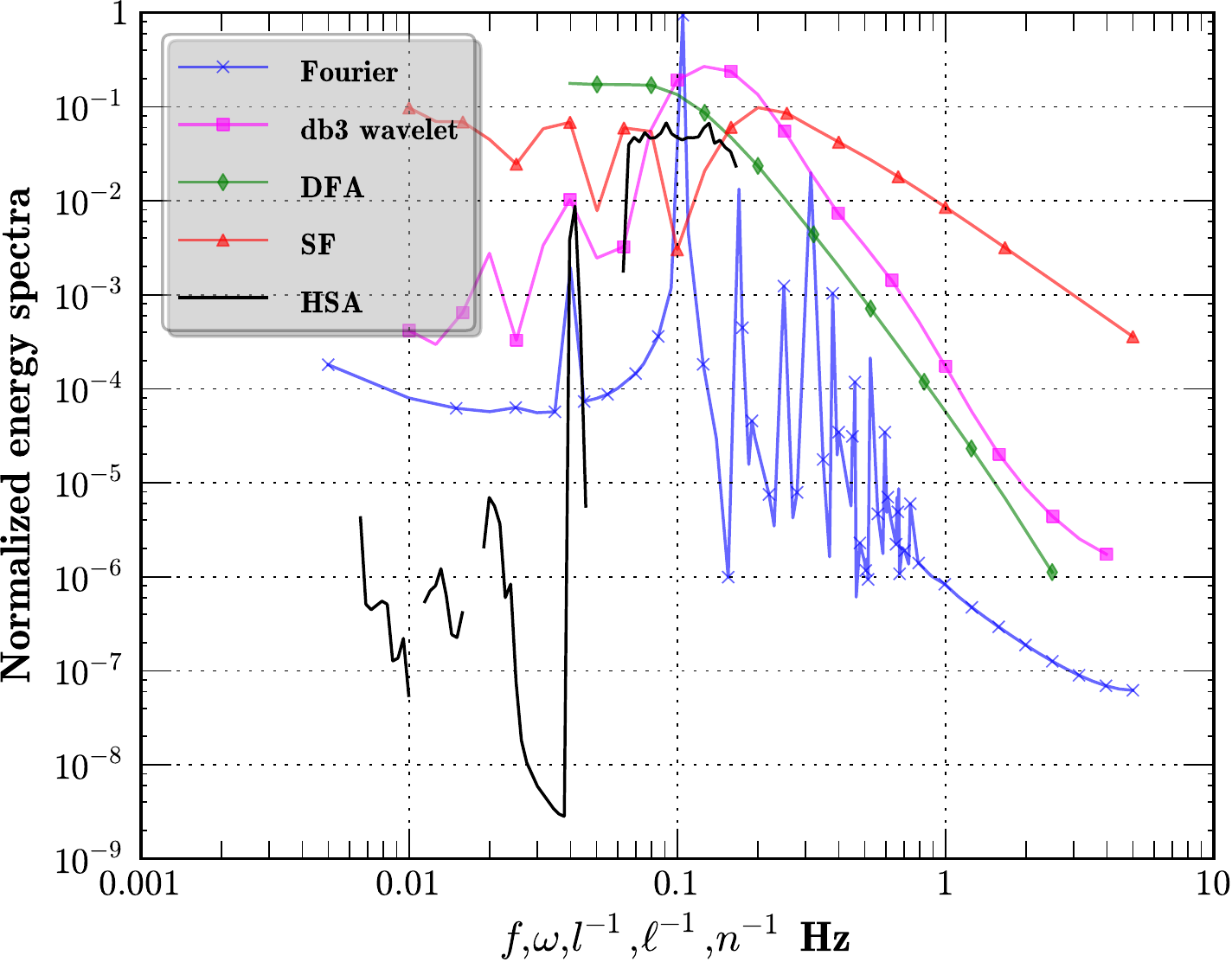}
\caption{(Color online) Energy spectra (the second-order statistical moments) provided by various
methods:  Fourier analysis ($\times$), continuous  wavelet transform with db3 wavelet ($\square$), first-order
DFA ($\diamondsuit$), SF ($\triangle$) and Hilbert
spectral analysis (solid
line. The wavelet, SF and DFA spectra have been converted into
spectral space
by taking $f=1/l$, $f=1/\ell$ and $f=1/n$,  respectively. Except for the Hilbert-based method,
all methods require 
high order harmonic components, which are not present  in the original signal,
see Figs. \ref{fig:Duffing} and \ref{fig:IF}, to represent  this  nonlinear process.
The high-order harmonics may lead to an artificial
energy flux from low frequencies (large scales) to  high frequencies (small scales) 
in spectral space.  
}\label{fig:spectra}
\end{figure}

We first consider  nonlinear effects by using the classical Duffing
equation, which reads 
\begin{equation}
\frac{\upd^2 x}{\upd t^2}+x(1+\epsilon x^2)=b\cos(\Omega t)
\end{equation}
 in which $\epsilon$ is a nonlinear parameter. 
It can be considered as a pendulum with forcing function $b\cos(\Omega t)$, in
which its pendulum
length varies with  the  angle.
Figure \ref{fig:Duffing} shows a fifth-order Runge-Kutta numerical solution
(thick solid
line) with $\epsilon=1$,
$b=0.1$, $\Omega=2\pi/25$, and $[x(0),x'(0)]=[1,1]$. The sampling frequency sets as 10\,Hz. For comparison, we also
show a pure sine wave  (thin solid line)   with the same mean frequency. One
can see that  the wave profile of  the  solution of  the  Duffing equation is
significantly deviating  from  a  sine wave. This deviation  is the result of
nonlinear
interaction, namely nonlinear distortion \cite{Huang1998EMD,Huang1999EMD}.
It is also clearly shown that there are no high-order harmonic components in the 
physical domain \cite{Huang2009PHD}.  Figure \ref{fig:IF} shows the first three
IMF modes obtained from
EMD decomposition and
the corresponding instantaneous
frequency $\omega$ from Hilbert spectral analysis. The instantaneous energy is
encoded
as  a   color.  The instantaneous frequency $\omega$ of the first IMF mode is varying
within  one period.  This corresponds to  the so-called
intrawave-frequency-modulation, which  is associated  with the nonlinear interactions
\cite{Huang1998EMD,Huang1999EMD,Huang2009PHD}. It  also clear shows that for a certain frequency, it may  not 
exist  clearly  all the time.

Figure \ref{fig:spectra} shows the normalized energy spectra (or the second-order statistical moments) provided by various methods:
 Fourier analysis ($\times$), continuous wavelet transform with db3  wavelet ($\square$), the first-order  DFA
($\diamondsuit$), SF ($\triangle$) and HSA (solid line). For display convenience, the wavelet,  DFA and
SF spectra have
been converted from physical domain into frequency domain by taking $f=1/l$,
$f=1/n$ and $f=1/\ell$,  respectively.  We emphasize here that  different wavelet families 
provide a similar spectral curve (not shown here). As pointed out  by \citet{Huang1998EMD}
   wavelet transform can be considered
 as an adjustable  window Fourier transform. Thus it inherits
 the  shortcomings  of the  Fourier transform. We observe that  except for the Hilbert
spectral analysis, all
methods
require high-order harmonic components to represent  this nonlinear process. High-order
 harmonic components are  not  present in the time series (see Figs. \ref{fig:Duffing} and \ref{fig:IF}). It is  thus a  requirement of
the method itself, not the physics 
\cite{Huang1998EMD,Huang2009PHD}.  This  is the main drawback of  traditional methods, in
which the basis is given \textit{a priori}. Therefore it is  inevitable that one requires high-order harmonic components to represent the
difference
between the  analyzed signal and the given basis. 
We argue here that high order harmonic components may lead to an artificial energy
flux from low frequencies (large scales) to high frequencies (small scales) in
spectral space.
Therefore, power law behavior, if it exists, may be contaminated by this
artificial energy flux.  We will show this point experimentally by analyzing a
temperature data set with strong ramp-cliff structures in Sec. \ref{sec:ps}.

%%%%Validation and Calibration%%%%%

\section{Validation and calibration}\label{sec:data}

%%%%%%%%%%%%%%%
%%%%%%%%%%%%%%%
\begin{figure*}[!htp]
\centering
%%----start of first subfigure----
\begin{minipage}{0.49\linewidth}
\label{fig:fbm:SF}
%% label for first subfigure
\includegraphics[width=0.95\linewidth]{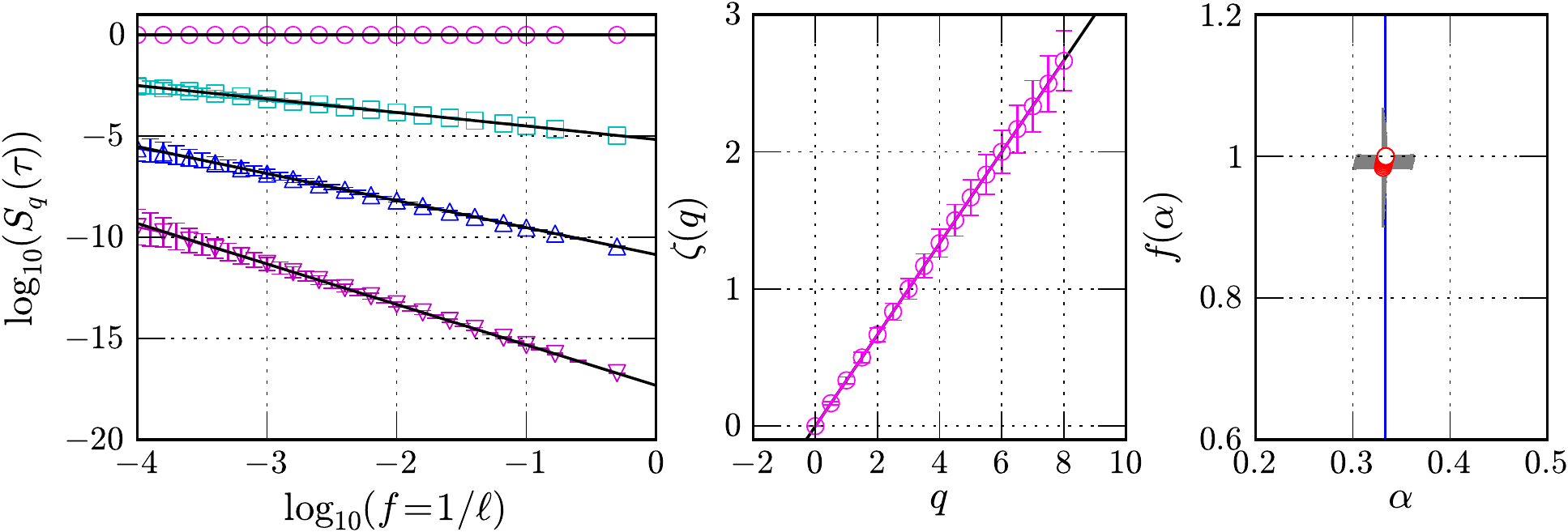}\\
(a)
\end{minipage}
%\hspace{.2cm}
%%----start of second subfigure----
\begin{minipage}{0.49\linewidth}
\label{fig:fbm:HSA}
%% label for second subfigure
\includegraphics[width=0.95\linewidth]{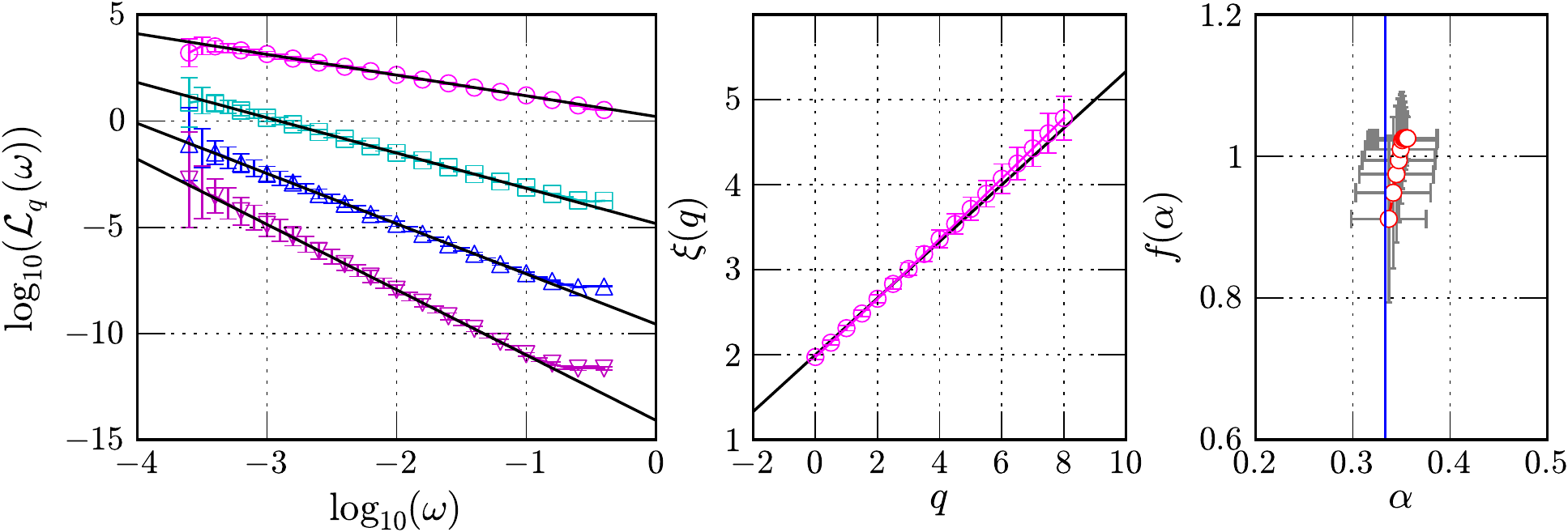}\\
(b)
\end{minipage}
\begin{minipage}{0.49\linewidth}
\label{fig:fbm:DFA}
%% label for third subfigure
\includegraphics[width=0.95\linewidth]{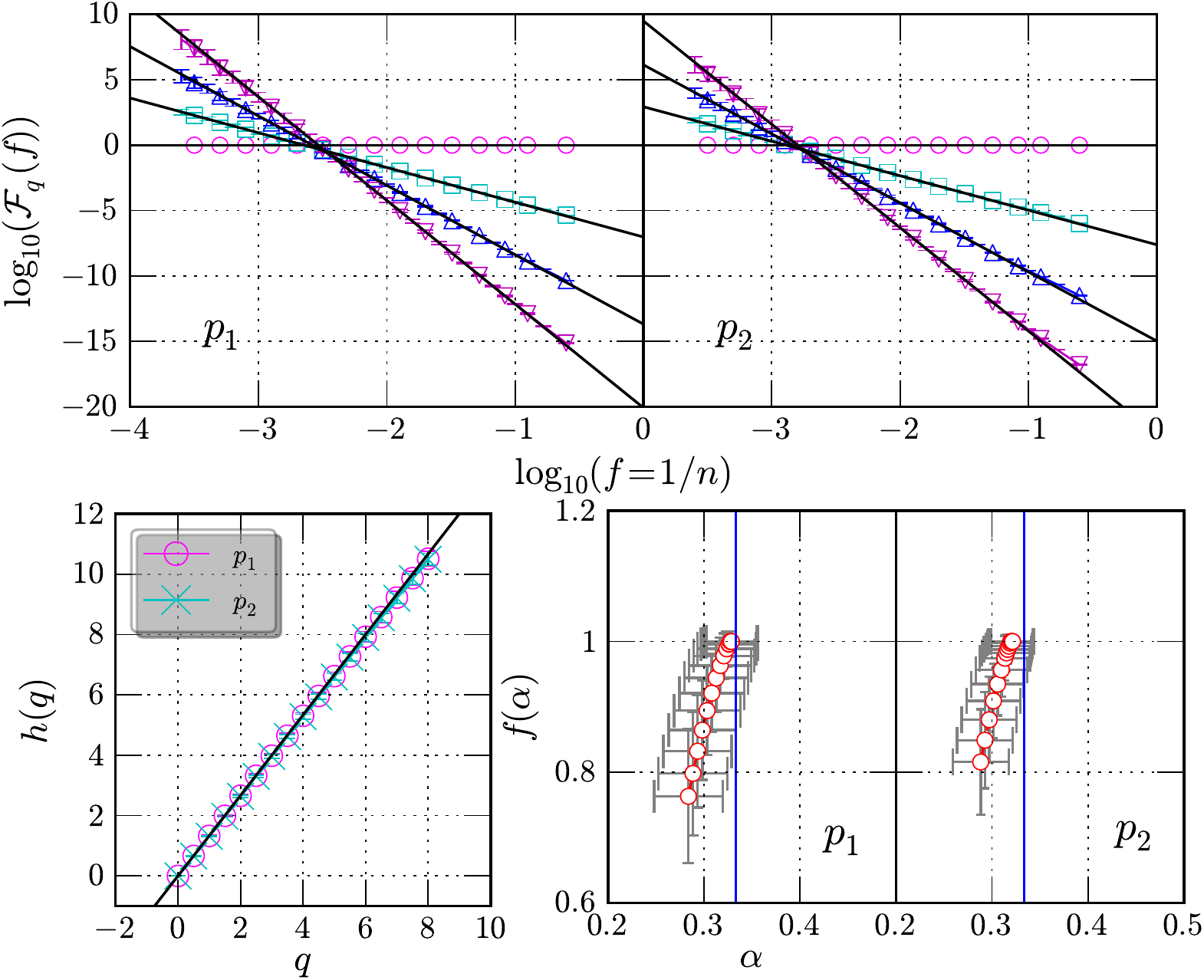}\\
(c)
\end{minipage}
%\hspace{1cm}
%%----start of fourth subfigure----
\begin{minipage}{0.49\linewidth}
\label{fig:fbm:wavelet}
%% label for fourth subfigure
\includegraphics[width=0.95\linewidth]{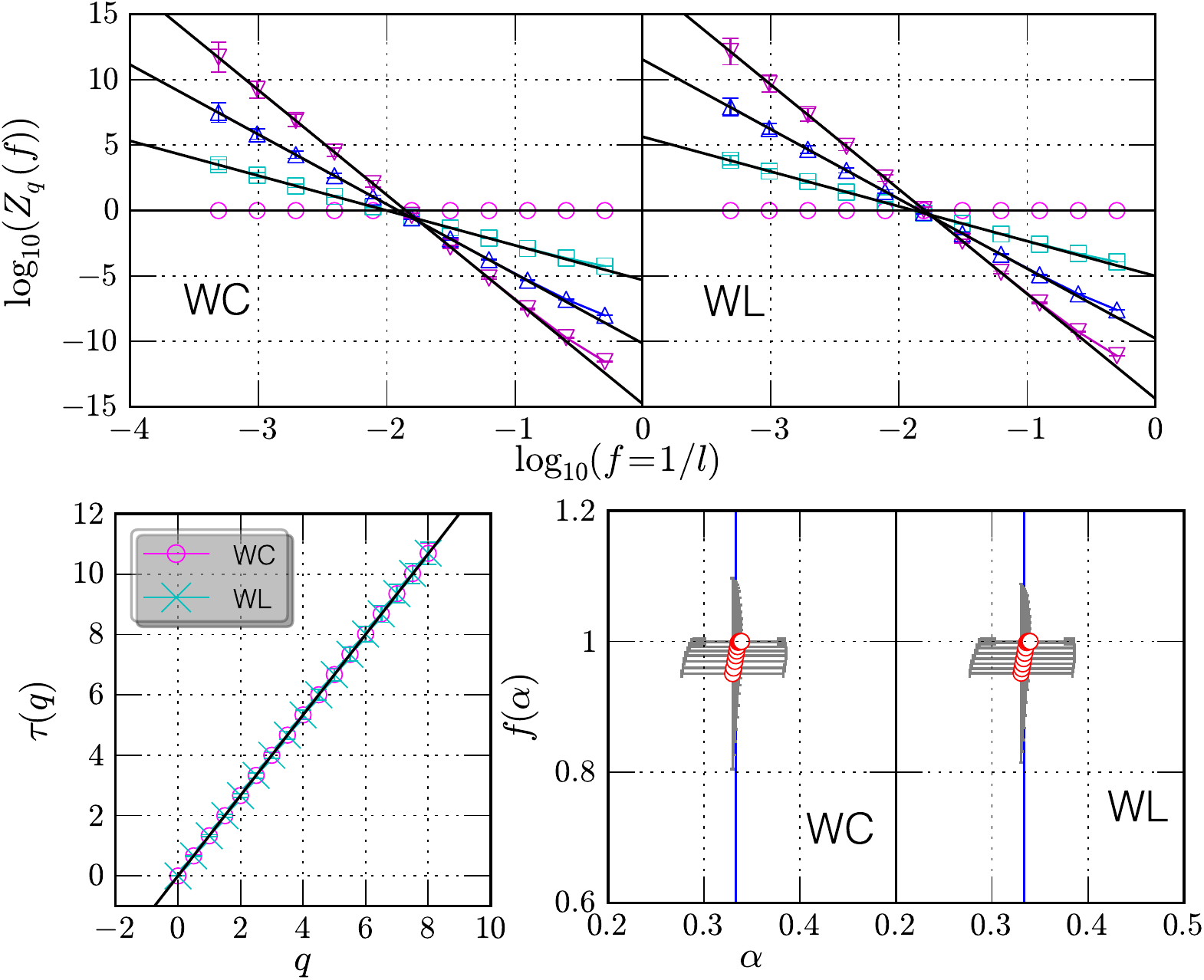}\\
(d)
\end{minipage}
\caption{(Color online) Analysis results of fBm with Hurst number $H=1/3$. (a)  structure functions. left:  $S_q(\ell)$ with $q=0$ ($\ocircle$), $q=2$ ($\square$), $q=4$
($\triangle$) and $q=6$ ($\triangledown$); middle: the corresponding scaling
exponents $\zeta(q)$ on the range $0\le q \le8$; right: the corresponding
singularity spectrum $h(\alpha)$.  (b) results of Hilbert spectral analysis.   (c) multifractal detrended fluctuation analysis (d) wavelet coefficients and wavelet leaders.  The symbols are the same as for structure functions.
Scaling exponents are estimated in  the range
 $-3<\log_{10}(f)<-1$. The statistical errors are estimated from a  total  of  500
realizations. }
\label{fig:fbm}
\end{figure*}

%%%%%%%%%%%%%%%
%%%%%%%%%%%%%%%
%%%%%%%%%%%%%%%

\begin{figure}[!htb]
\centering
 \includegraphics[width=0.85\linewidth]{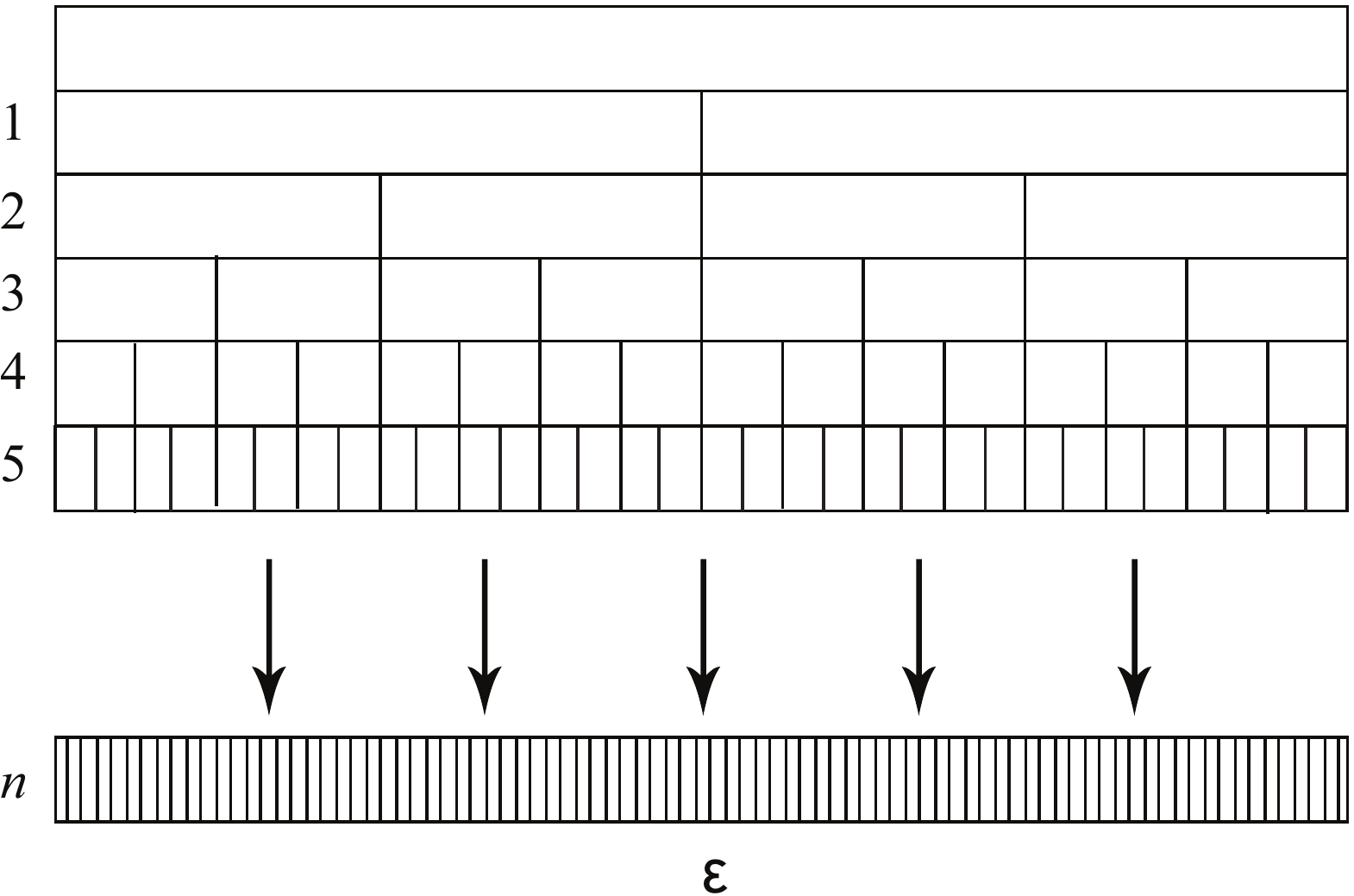}
  \caption{Illustration of the discrete cascade process. Each step is associated
with  a scale ratio of 2. 
After $n$ steps, the total scale ratio is $2^n$. }
  \label{fig:cascade}
\end{figure}

\begin{figure}[!htb]
\centering
 \includegraphics[width=0.85\linewidth]{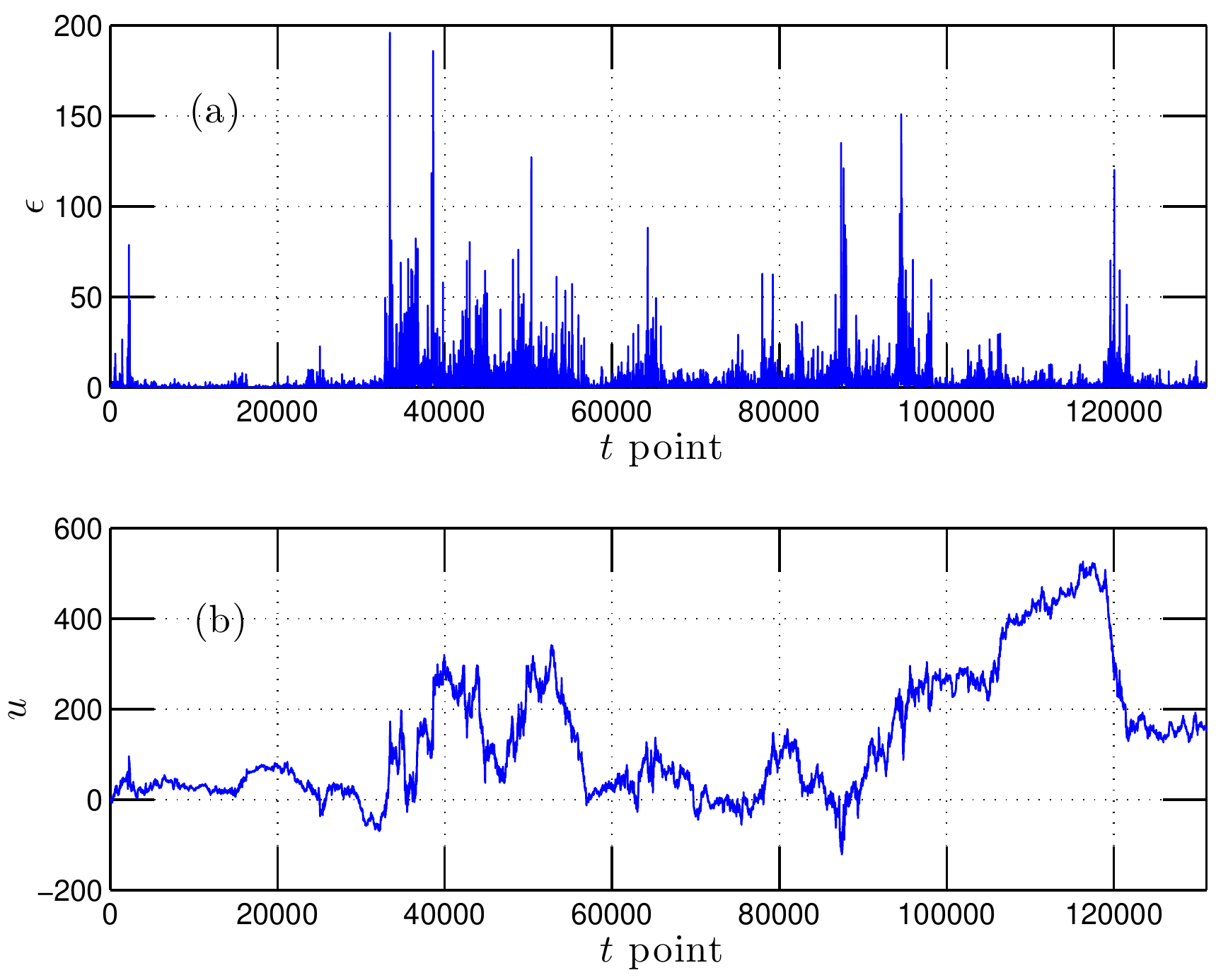}
  \caption{A sample for one  realization of length $2^{17}$ points with $\mu=0.15$,
  (a) the multifractal measure, and (b) the constructed multifractal nonstationary
process.  }\label{fig:portion}
\end{figure}

%%%%%%%%%%%%%%%
%%%%%%%%%%%%%%%
%%%%%%%%%%%%%%%
\begin{figure*}[!htp]
\centering
%%----start of first subfigure----
\begin{minipage}{0.49\linewidth}
\label{fig:mfbm:SF}
%% label for first subfigure
\includegraphics[width=0.95\linewidth]{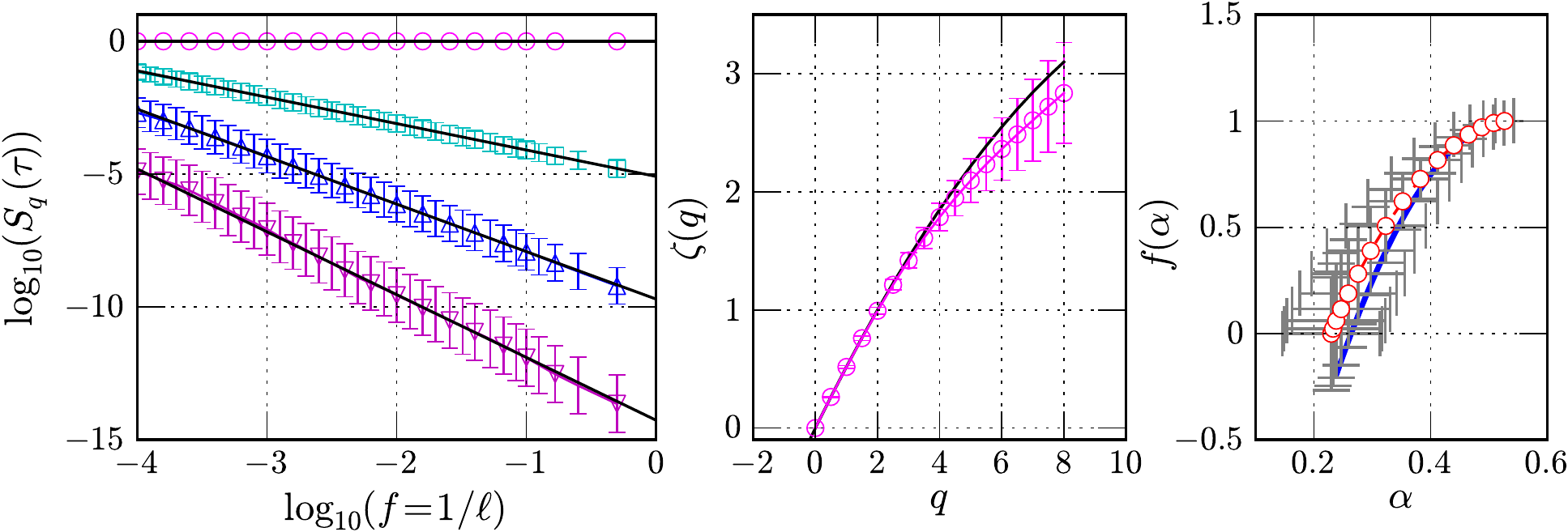}\\
(a)
\end{minipage}
 %%----start of second subfigure----
\begin{minipage}{0.49\linewidth}
\label{fig:mfbm:HSA}
%% label for second subfigure
\includegraphics[width=0.95\linewidth]{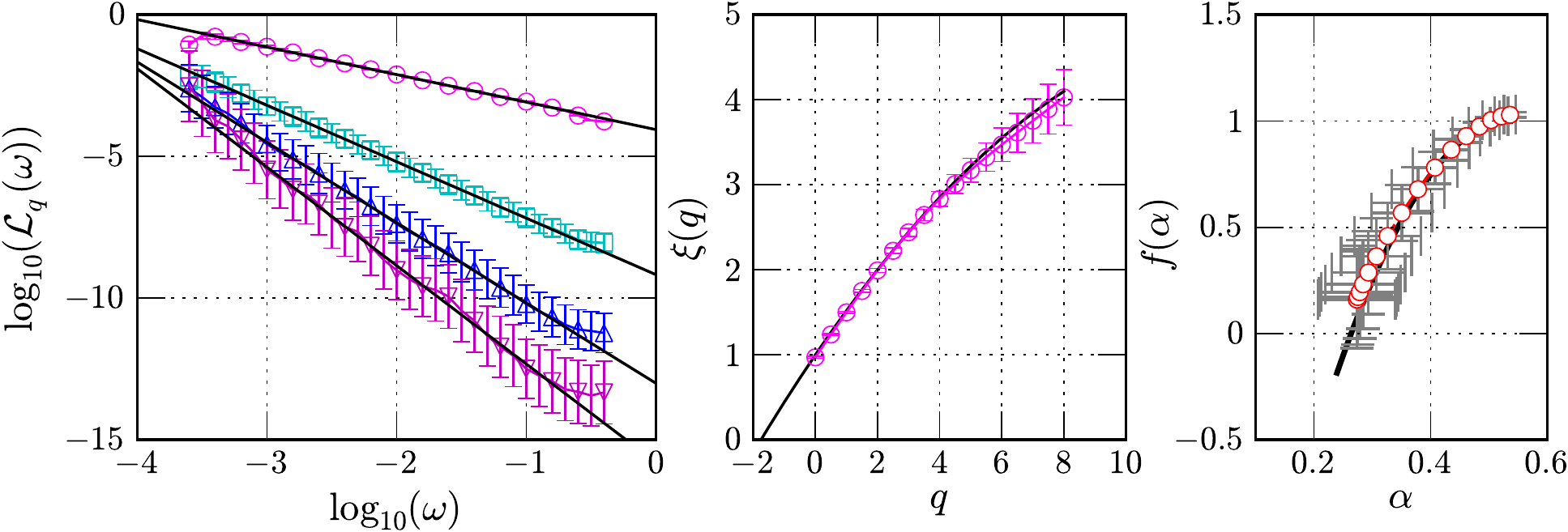}\\
(b)
\end{minipage}
\begin{minipage}{0.49\linewidth}
\label{fig:mfbm:DFA}
%% label for third subfigure
\includegraphics[width=0.95\linewidth]{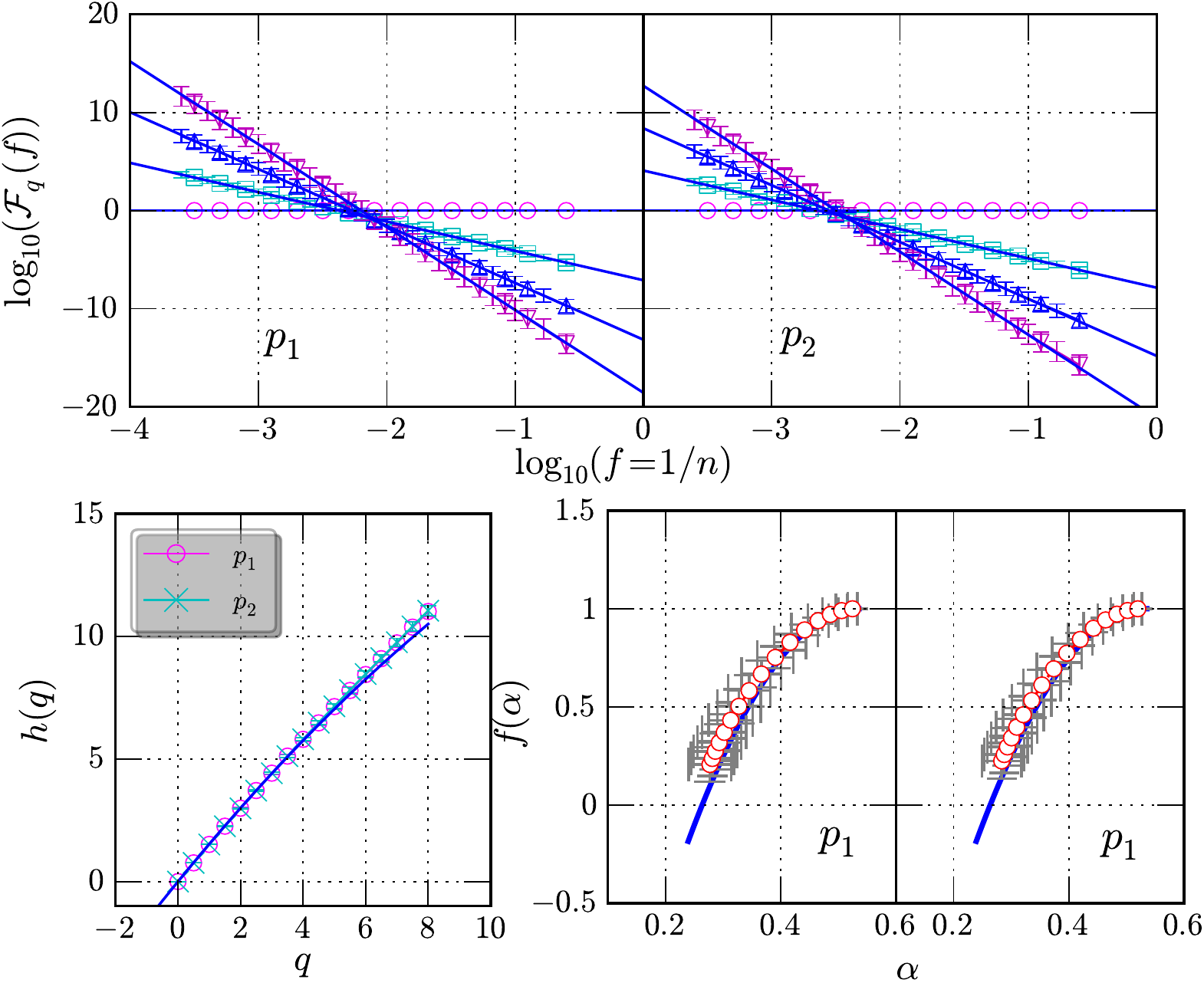}\\
(c)
\end{minipage}
 %%----start of fourth subfigure----
\begin{minipage}{0.49\linewidth}
\label{fig:mfbm:wavelet}
%% label for fourth subfigure
\includegraphics[width=0.95\linewidth]{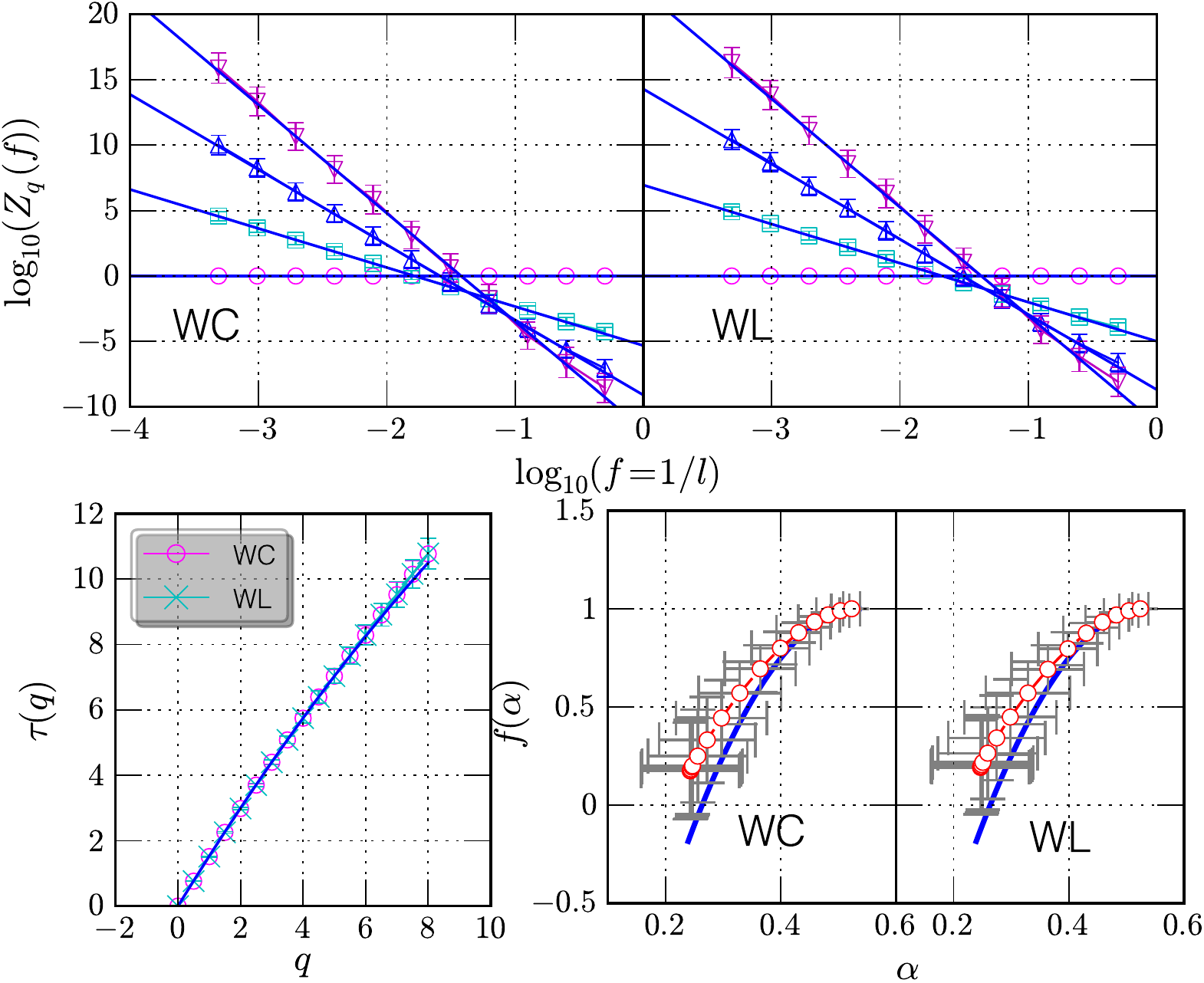}\\
(d)
\end{minipage}
\caption{(Color online) Multifractal random walk with $\mu=0.15$. (a) Structure function , (b) Hilbert spectral analysis,  (c) multifractal detrended fluctuation analysis and  (d) wavelet coefficients and wavelet leaders.
The symbols are the same as in Fig. \ref{fig:fbm}. The statistical error bars
are estimated from the total 100 realizations. }
\label{fig:mfbm:one}
%% label for first part
\end{figure*}

%%%%%%%%%%%%%%%
%%%%%%%%%%%%%%%
%%%%%%%%%%%%%%%

In this section, we will validate the Hilbert-based method by performing
a comparison study of simulated fBm 
with Hurst number $H=1/3$ and synthesized  multifractal random walk with  an 
intermittent
parameter $\mu=0.15$. For comparison convenience, 
spectral curves (or the $q$th-order statistical moment) provided by  SFs,
MFDFA and wavelet are converted from the physical domain into the spectral
domain by taking $f=1/\ell$, $f=1/n$, and $f=1/l$,  respectively.
 The corresponding scaling exponents are estimated on the range
$0.001<f<0.1$ (we set here the sampling frequency  as 1). Wavelet transform  is performed by 
using the db3 wavelet. Due to the limitation of the SF analysis
and the HSA, we only consider here the non-negative $q$th-order 
 moment, $q\ge 0$, the left part of the singularity spectrum.

\subsection{Fractional Brownian motion}

We have shown in previous works that the arbitrary-order HSA can be 
applied to the fractional Brownian motion \cite{Huang2008EPL,Huang2008TSI}. Here
we
briefly recall these results. FBm is a
Gaussian self-similar process with a normal distribution increment, which is
characterized by $H$,  the  Hurst number $0<H<1$
\cite{Beran1994,Rogers1997,Doukhan2003,Gardiner2004}.
Note that the singularity spectra for the above mentioned methods are 

\begin{subequations}
\begin{equation}
\alpha= \zeta'(q), \quad f(\alpha)=\min_{q}\{ \alpha q-\zeta(q) +1\}
\end{equation}
for SFs, and
\begin{equation}
\alpha=\xi'(q),\quad f(\alpha)=\min_{q}\{ \alpha q-\xi(q)+2 \}
\end{equation}
 for the Hilbert-based method, and
\begin{equation}
\alpha=h'(q)-1,\quad f(\alpha)=\min_{q}\{(\alpha+1)q-h(q)+1 \}
\end{equation}
for DFA, and
\begin{equation}
\alpha=\tau'(q)-1,\quad f(\alpha)=\min_q\{ (\alpha+1)q-\tau(q)+1\}
\end{equation}
 for WC and WL, respectively.
\end{subequations}
Ideally, we should have $\alpha=H$ and $f(\alpha)=1$. 

We performed 500 realizations each of length  $2^{14}$ data points by
  applying   a Fourier-based Wood-Chan algorithm  \cite{Wood1994}  with $H=1/3$, which
corresponds to the Hurst number
of turbulent velocity.   
We apply the above mentioned methods to each
realization  of the data series. The final spectra and statistical errors are then estimated
from  these 500 realizations. 
Figure  \ref{fig:fbm} show results for (a) SF:   (left)
$S_q(\ell)$ with $q=0$ ($\ocircle$), $q=2$ ($\square$), $q=4$ ($\triangle$)
and $q=6$ ($\triangledown$), (middle) the corresponding scaling exponents $\zeta(q)$
on the range $0\le q \le 8$, (right) the corresponding singularity spectrum
$f(\alpha)$, (b) HSA, (c) DFA, and (d) wavelet,
respectively. The symbols are the same as the SF symbols. 
Graphically, all methods
provide comparable estimation of $f(\alpha)$. 
However, we note that the Hilbert-based method slightly  overestimates $\xi(q)$ when $q>6$.
Additionally both the first- and second-order DFA provide   slight  underestimation of $h(q)$ and 
seem to  predict a systematic underestimation of the Hurst number $H$.
The WC and WL provide almost the same estimation for this simple monofractal
process. It seems that they provide a better estimation than Hilbert and DFA methods.
This result is  not in full agreement  with \citet{Oswicecimka2006PRE}, who stated
that the MFDFA provides a better estimation of $H$ than WTMM.

 The above results show that all methods provide comparable prediction of
singularity spectra for fBm with $H=1/3$ . However, it seems that SF and wavelet
based methods provide a better estimation.

\subsection{Multifractal simulation}

We show now that the new method applies to multifractal time series.
First, let us consider  a multiplicative discrete  {cascade} process to simulate a multifractal  measure $\epsilon(x)$.
Figure~\ref{fig:cascade} illustrates the cascade process algorithm.  The larger scale corresponds to a unique cell of size $L=\ell_0 \lambda_1^n$, where $\ell_0$ is a fixed 
scale and $\lambda_1>1$ is   a  dimensional scale ratio. For discrete models, this ratio is often taken as $\lambda_1=2$. The models being discrete, the next scale involved 
corresponds to $\lambda_1$ cells, each of size $L/\lambda_1=\ell_0 \lambda_1^{n-1}$. This is iterated and at step $p$ ($1\le p \le n$) there are $\lambda_1^p$ cells, 
each of size $L/\lambda_1^p=\ell_0 \lambda_1^{n-p}$. There are $n$ cascade steps, and at step $n$ there are $\lambda_1^{n}$ cells, each of size $\ell_0$, which is 
the smallest scale of the cascade. To reach this scale,  all intermediate scales have been involved. Finally, at each point the multifractal measure writes as the product of 
$n$ cascade random variables
\begin{equation}
  \epsilon(x)=\prod_{p=1}^n W_{p,x}
\end{equation}
where $W_{p,x}$ is the random variable corresponding to position $x$ and level $p$ in the cascade~\cite{Schmitt2003}.
Following multifractal random
walk ideas \cite{Bacry2001,Muzy2002}, we generate a nonstationary multifractal time series as
\begin{equation}
 u(x)=\int_{0}^x {\epsilon(x')^{1/2}} \upd B(x')\label{eq:multitime}
\end{equation}
where $B(x)$ is Brownian motion. Taking  lognormal statistic for $\epsilon$,
the scaling exponent $\zeta(q)$ such as $\langle \vert\Delta u_{\tau}(t)\vert^q\rangle\sim
\tau^{\zeta(q)}$ can be shown to  be written as
\begin{equation}
\zeta(q)=\frac{q}{2}-\frac{\mu}{2}(\frac{q^2}{4}-\frac{q}{2})\label{eq:lognormal}
\end{equation}
where $\mu$ is the  intermittency parameter ($0\le\mu\le1$) characterizing
the lognormal multifractal cascade.

Synthetic multifractal time series are generated following Eq.\eqref{eq:multitime}. For each realization, we choose  $n=17$ levels, corresponding to data sets with data length  $131,072$ points each. A
sample for one realization is shown in Fig.~\ref{fig:portion} (a) for the multifractal
measure and (b) for the nonstationary multifractal time series with $\mu=0.15$. We perform
100 realizations with intermittent parameter $\mu=0.15$. Except  for
 the structure
functions, we apply all  methods to each realization by dividing one realization
into eight subsets with $2^{14}$ data points each. The spectra for each realization
are averaged over these eight subsets. The final spectra and error bars are respectively ensemble average and standard deviation estimated from these 100 realizations.

Figure \ref{fig:mfbm:one} shows the results of (a) SF, (b)  HSA,
(c) MFDFA and
(d) WC and WL, respectively. The symbols are the same as in Fig. \ref{fig:fbm}.  The theoretical scaling exponents and the corresponding singularity spectrum $f(\alpha)$  on the range $0<q<8$ are shown as a solid line in the corresponding sub figures.
 We see that  SFs underestimate $\zeta(q)$  when $q>4$. The corresponding estimated singularity spectrum $f(\alpha)$ deviates from the theoretical line when $\alpha<0.4$, corresponding to $q>2.5$. It also has the largest  statistical error.
Hilbert methodology slightly underestimates $\xi(q)$ when $q>5$. It provides a better estimation of scaling exponents and  
$f(\alpha)$ than SFs.   MFDFA provides the smallest statistical errors for 
 spectral  curves $\mathcal{F}_q(n)$, scaling 
exponents $h(q)$ and singularity spectrum $f(\alpha)$. However, it still  slightly  overestimates $h(q)$ when $q>6$.  We  note that  the first- and second-order DFA provide an equivalent result.   WC and WL predict almost the same spectral curves, scaling exponents $\tau(q)$ and singularity spectrum $f(\alpha)$.  The corresponding singularity spectrum  significantly deviates from the theoretical curve. We also note that none of  these methods  recover the whole theoretical line on the range $0<q<8$.

 \section{Passive scalar turbulence with ramp-cliff structures}\label{sec:ps}

\begin{figure}[!htb]
\centering
 \includegraphics[width=0.85\linewidth]{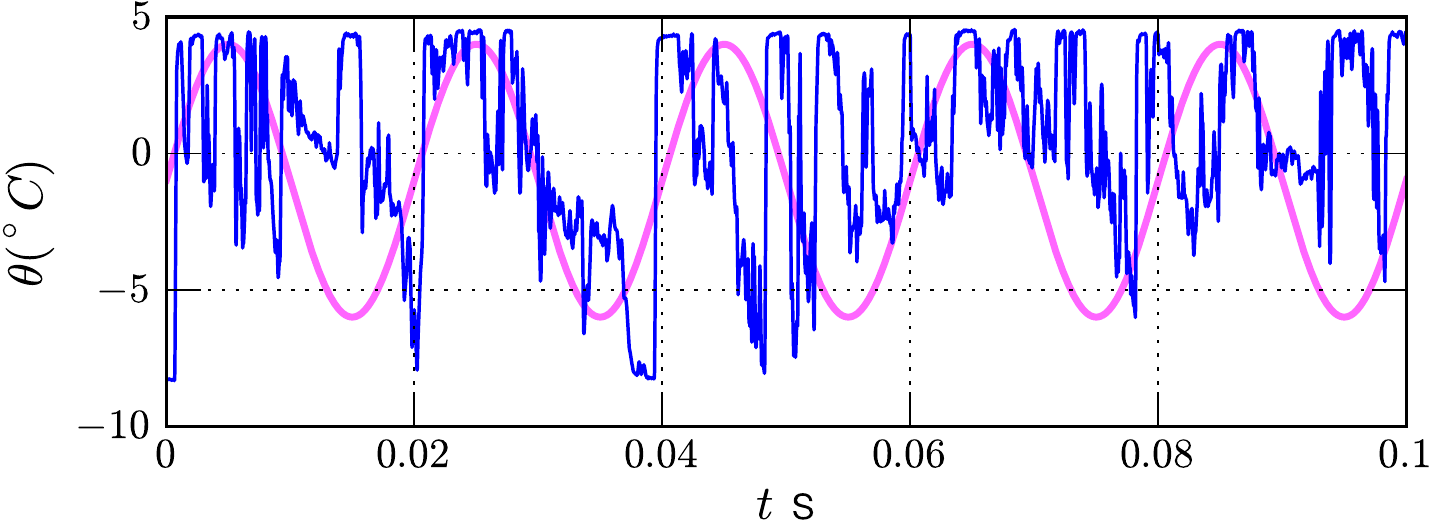}
  \caption{(Color online) A 0.1\,s portion of temperature  data showing strong ramp-cliff
structures. For comparison,  a  pure sine wave is also shown as  a  thick solid
line.  Note  that the ramp-cliff structure is significantly different  from a  sine
wave, which may cause serious  artificial energy flux in Fourier spectral
space.}
  \label{fig:portionT}
\end{figure}

\begin{figure}[!htb]
\centering
 \includegraphics[width=0.85\linewidth]{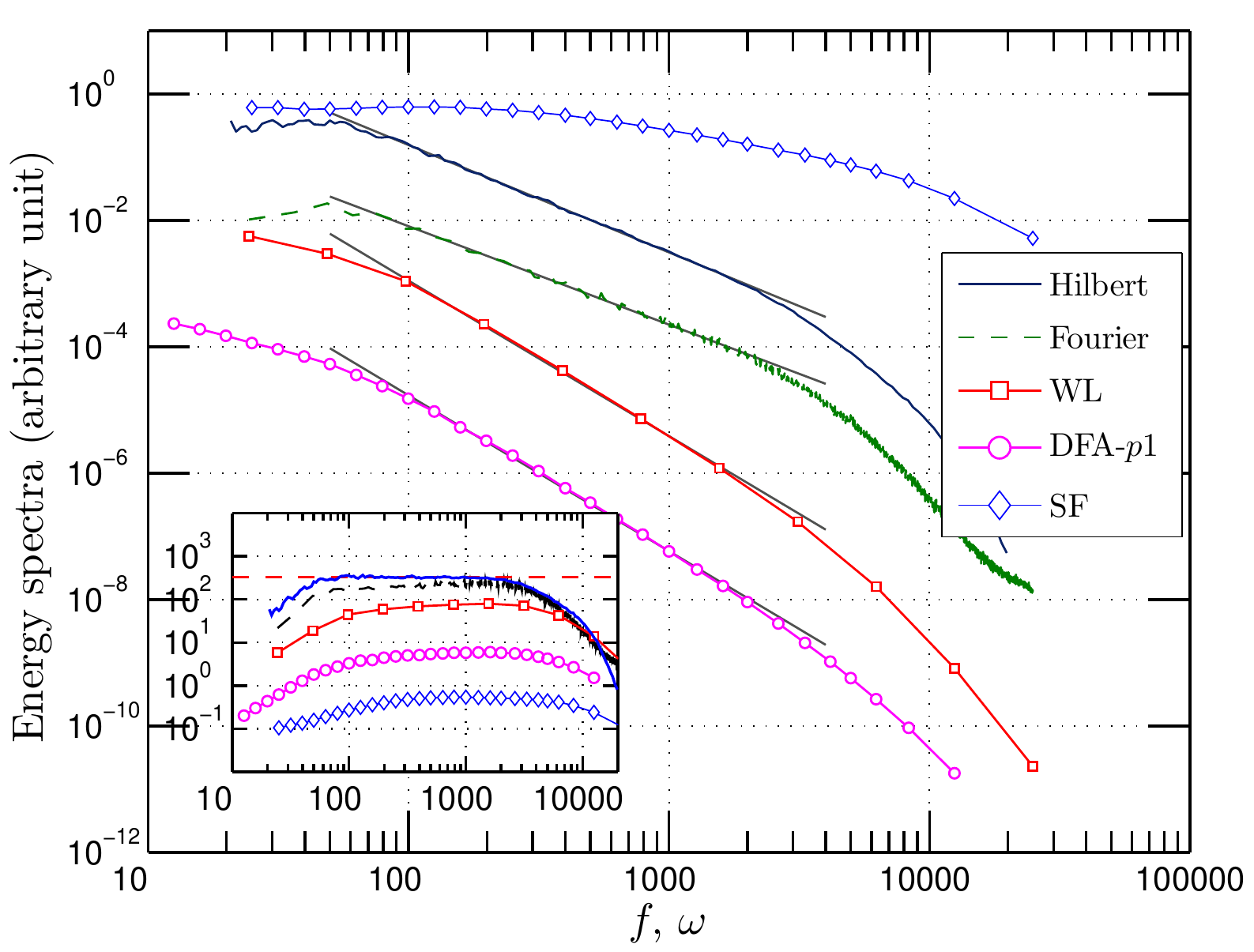}
  \caption{(Color online) Energy spectra (or the second-order statistical moment) provided by several methods.  
  The inset shows the
compensated spectra by multiplying the   result by  $f^{5/3}$ for Hilbert and
Fourier, $f^{8/3}$ for DFA and wavelet, and $f^{2/3}$ for SFs,
respectively.
 For clarity,  the curves
have been  vertically  shifted. Both Fourier and Hilbert methods predict   a clear
power
law  on the range  $80<\omega<2000\,$Hz. Due to the presence of ramp-cliff
structures, the SF analysis fails to capture the power law behavior and
DFA and wavelet predict a   short  inertial range on the range $100<f<1000
\,$Hz. The
corresponding scaling exponents are $\beta_{\theta}=1.56$ for Fourier,
$\xi_{\theta}(2)=1.70$ for
Hilbert,
$\tau_{\theta}(2)=2.46$ for WL with db3 wavelet, and
$h_{\theta}(2)=2.47$ for the
first-order DFA, respectively. }
  \label{fig:passivespectra}
\end{figure}

\begin{figure*}[htp]
%\ContinuedFloat
\centering
\begin{minipage}{0.65\linewidth}
\label{fig:passive:HSA}
\includegraphics[width=0.95\linewidth]{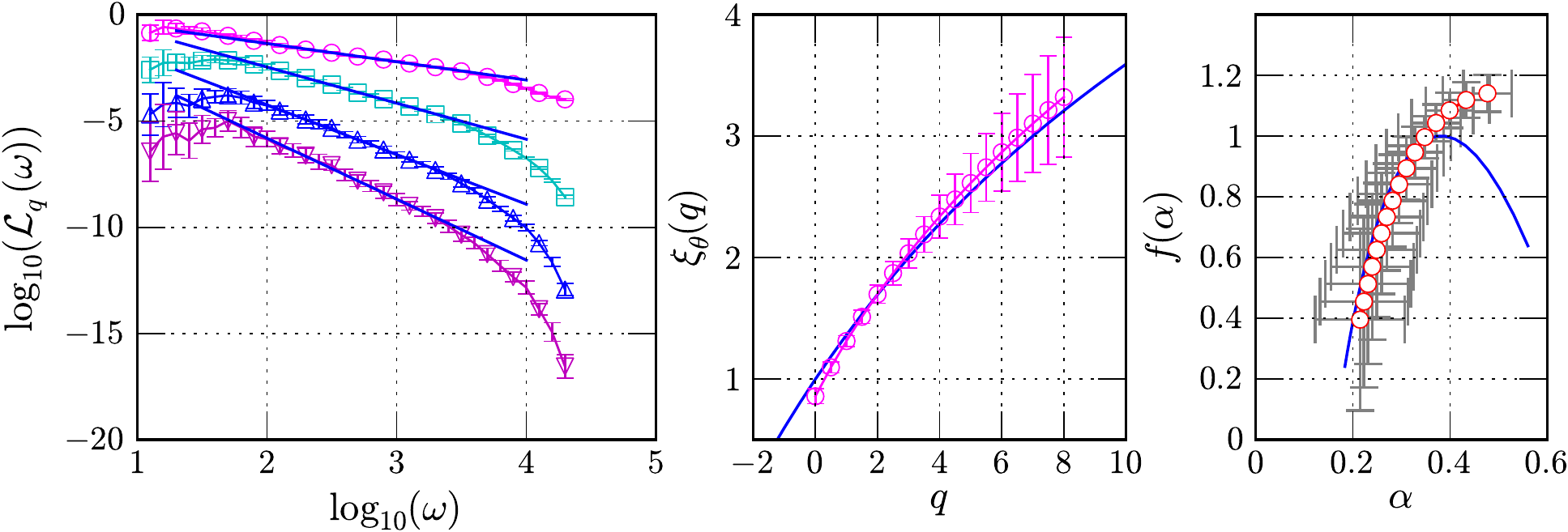}\\
(a)
\end{minipage}\\
\begin{minipage}{0.49\linewidth}
\label{fig:passive:DFA}
%% label for third subfigure
\includegraphics[width=0.95\linewidth]{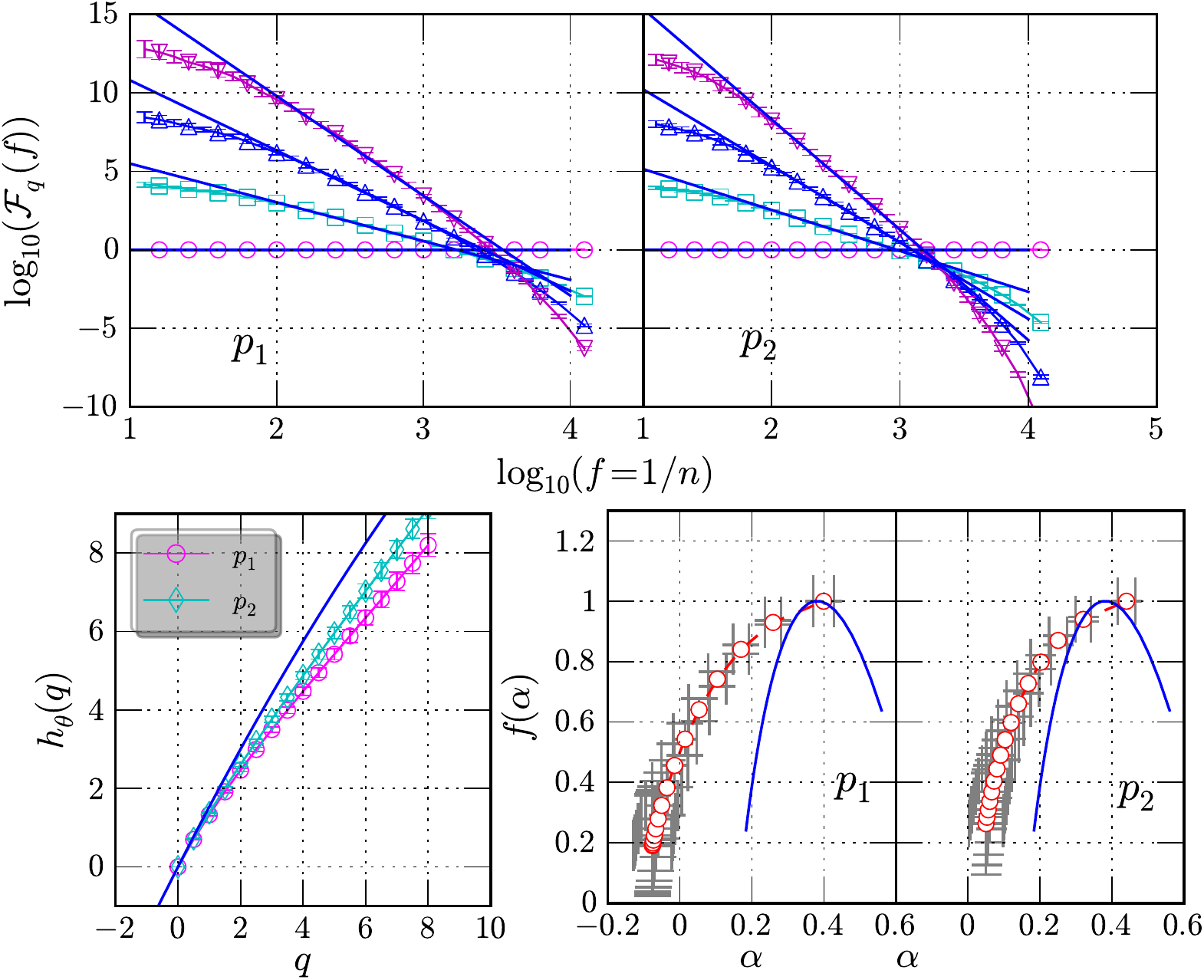}\\
(b)
\end{minipage}
%%----start of fourth subfigure----
\begin{minipage}{0.49\linewidth}
\label{fig:passive:wavelet}
%% label for fourth subfigure
\includegraphics[width=0.95\linewidth]{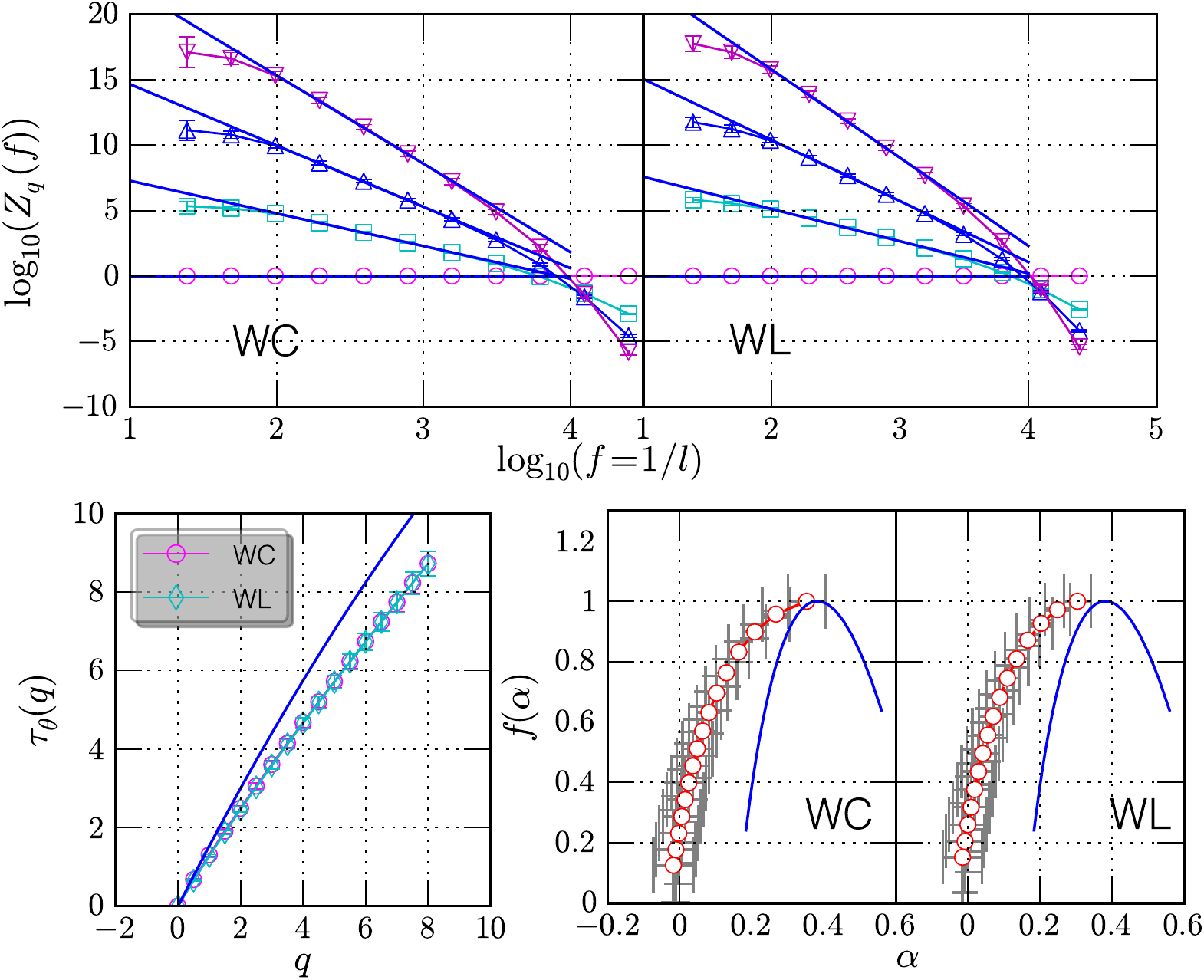}\\
(c)
\end{minipage}
\caption{(Color online) Results for passive scalar (temperature) with strong ramp-cliff (a) Hilbert spectral analysis, (b) multifractal detrended fluctuation analysis, and (c) wavelet coefficients and wavelet leaders. The symbols are the same as in Fig. \ref{fig:fbm}. }
\label{fig:passive:two}
%% label for second part
\end{figure*}

 We now  apply the above-mentioned methods to a real time data set, a temperature time series as a 
 turbulent  passive scalar. 
The data  are  obtained  from   a jet experiment performed at  Joseph Fourier University  Grenoble, France.  The bulk Reynolds number is about $Re\simeq 60000$. The corresponding Taylor\rq{}s microscale Reynolds number 
is about $Re_{\lambda}\simeq 250$. The initial
temperature of the two flows are $T_{J}=27.8 ^{\circ} C$ and $T=14.8 ^{\circ} C$. The
measurement location is in the mixing layer and close to the nozzle of the jet. 
The sampling frequency is $50\,$kHz. The total data 
length is 10\,s, corresponding to 500,000 data points. Figure \ref{fig:portionT} shows a 0.1\,s portion
temperature data,   illustrating strong ramp-cliff structures. For comparison, a pure sine wave is also shown. 
Obviously, the so-called ramp-cliff structure is  a  large-scale 
structure with a very sharp interface \cite{Sreenivasan1991,Shraiman2000,Warhaft2000,Celani2000}.  We  note that 
the profile of ramp-cliff structures is  significantly
deviating  from  a  sine wave.  Thus  for  the Fourier-based methodologies,
 it  is inevitable that one  requires high-order harmonic components to represent their difference, in which 
the underlying idea 
is a linear asymptotic approximation \cite{Cohen1995,Flandrin1998,Huang1998EMD}.  This linear asymptotic 
approximation process thus leads  to  an artificial energy flux from low 
frequencies (large scales) to higher frequencies (small scales). It means that the Fourier-based  
spectrum may be contaminated by this artificial energy flux. As another direct consequence, the artificial 
redistribution of the energy  will lead to an unreal correlation if we consider cross-correlation between two 
scales \cite{Huang2011PRL}.

The original time series is divided into 122 non-overlapping segments with $2^{12}$ data points each.
The finally spectra and statistical errors (the standard deviation) are then estimated from these 122 
realizations.
Figure \ref{fig:passivespectra} shows the energy spectra (or the second order statistical moments)  provided by  
HSA (solid line), Fourier transform (dashed line),  WL ($\square$), the first-order DFA 
($\ocircle$) and SF ($\diamondsuit$), respectively.   The inset shows the 
corresponding compensated spectra by multiplying a Kolmogorov- Obukhov-Corrsin 
\cite{Kolmogorov1941,Obukhov1949,Corrsin1951} nonintermittent scaling exponent $5/3$ for Hilbert spectrum, 
Fourier power 
spectrum,  ${8/3}$  for WL and DFA, and ${2/3}$ for SF, respectively. Except for the SF, all  methods 
display a clear power law on the  range $80<f<2000\,$Hz or $100<f<1000\,$Hz, a more than one decade inertial 
range.  The corresponding scaling exponents are $\xi_{\theta}(2)\simeq 
1.70$ for Hilbert, $\beta_{\theta}\simeq 1.56$ for Fourier, $\tau_{\theta}(2)\simeq 2.46$ for WL
 and $h_{\theta} (2)\simeq 2.47$ for DFA, 
respectively obtained by using a least square fitting algorithm. We note that only the Hilbert based scaling 
exponent $\xi_{\theta}(2)$ is close to the corresponding nonintermittent scaling exponent  $\xi_{\theta}(2)=5/3$ 
\cite{Huang2010PRE}. It is also
 comparable with the scaling exponent of longitudinal velocity in fully developed turbulence 
 \cite{She1994,Frisch1995,Water1999}.
 Due to the presence of  strong ramp-cliff structures, the SF fails to detect the correct 
 scaling behavior. The influence of large energetic structures on SF has been  studied in detail by 
 \citet{Huang2010PRE} and \citet{Huang2009PHD}.
 It is interesting to note that DFA and WL provide almost the same scaling exponent, which indicates that the 
 ramp-cliff structure may have the same influence on  them.
  We believe that there exists an artificial energy flux as we discussed  above in both Fourier and DFA and WL 
  spectra. Thus they may underestimate the scaling exponents \cite{Huang2010PRE}. 
 
 Figure \ref{fig:passive:two}  shows  the analysis results of (a)   Hilbert spectral analysis, (b) MFDFA, 
 and (c) wavelet transform, respectively.   The symbols are the same as in Fig. \ref{fig:fbm}. For 
 comparison, the  lognormal  model of longitudinal velocity \cite{Schmitt2006} is shown as a solid line in 
the  subfigures of scaling exponents and singularity spectra. Due to the failure  of  SF analysis, we do not present 
  it  here (see Ref. \cite{Huang2009PHD}).  Graphically, these three methodologies predict   power law 
 spectra with  small statistical  error.
 The corresponding scaling exponents are estimated on the range $80<f<2000\,$Hz or $100<f<1000\,$Hz. It is found 
 that  the corresponding scaling exponents $\xi_{\theta}(q)$ and singularity spectrum $f(\alpha)$ are close to 
 the  lognormal fitting  model, indicating a less intermittent passive scalar turbulence field than what was believed before \cite{Huang2010PRE}.   MFDFA and wavelets provide comparable statistical errors and
 singularity spectra.  Their scaling exponents and singularity spectra  significantly deviate from 
  lognormal  model, which is usually  considered  as   evidence that the passive scalar turbulence 
 field  is much  more  intermittent than  the   velocity field 
 \cite{Sreenivasan1991,Shraiman2000,Warhaft2000,Celani2000}.  We note that the first- and second-order MFDFA 
 provide  different scaling exponents and singularity spectra, which may be associated with the different abilities 
 of different order polynomials \cite{Chen2002PRE,Hu2001}.

 As we already mentioned previously,   the wavelet and DFA spectra are strongly influenced by nonlinear large
 scale structures   (e.g. ramp-cliff structures in passive scalar turbulence). Their scaling exponents are thus 
 contaminated by high-order harmonics. In other words, the statistical property  of small scales
 is contaminated by nonlinear large-scale structures. We believe here that the scaling exponents and singularity 
 spectrum provided by them are not correct.  
 Since the HSA has a very local ability in both physical and 
 spectral spaces, together with the ability of intrawave-frequency-modulation for nonlinear processes, the 
 effect of ramp-cliff structures is   constrained in the amplitude-frequency space. Therefore, the HSA method 
  may provide  a more correct scaling exponent and singularity spectrum. We  note that for DFA and the wavelet method, the large deviation from a lognormal spectrum may be interpreted as a  shift 
  problem for moment-based methods when the translational invariance is broken\cite{Pont2006PRE,Turiel2006JCP}. It seems that the Hilbert-based method can automatically correct this problem.
   We also underline here that the 
 Reynolds number of the present passive scalar data set is about $Re_{\lambda}\simeq 250$. 
 Thus the strong ramp-cliff structure may be recognized as an effect of the finite Reynolds number. We will address 
 this issue  elsewhere.
 
\section{Conclusion}\label{sec:conclusion}

In summary,  we introduced in this paper a new method, namely arbitrary-order Hilbert spectral analysis, to 
characterize scale-invariant properties directly in the amplitude-frequency space 
\cite{Huang2008EPL,Huang2008TSI,Huang2009PHD}.  It is an extended version of Hilbert-Huang transform 
\cite{Huang1998EMD,Huang1999EMD,Huang2005EMDa}. The main advantage of the Hilbert-based methodology is its  fully 
adaptive \cite{Flandrin2004EMDa} and very local ability both in spectral  and 
physical domains 
\cite{Huang1998EMD,Huang1999EMD}.  Thus, it is not necessary to require high-order harmonics to represent nonlinear 
and nonstationary processes, 
which is usually required by conventional Fourier-based methods, such as Fourier transform, wavelet transform, 
etc. We illustrated  the nonlinear effect by using  the  Duffing equation. It is found that not only
Fourier-based methods, but also SF analysis
 and DFA are influenced by nonlinear processes. It is also found that the 
HSA can constrain the high-order harmonics by using the intrawave-frequency-modulation mechanism for the 
nonlinear distortion \cite{Huang1998EMD,Huang1999EMD,Huang2005EMDa}.

We then performed a comparison study of the Hilbert-based methodology  with SF analysis, MFDFA, and WL,
 by analyzing  fBm simulations with Hurst number $H=1/3$ and a synthesized multifractal  lognormal  random 
 walk with intermittent parameter $\mu=0.15$, respectively.  For the former simulation, we considered  the 
 scaling exponents and singularity spectrum on the range $0<q<8$. It  was  found that all methods 
 provide comparable  scaling exponents and  singularity spectra. 
 For the latter synthesized multifractal random walk data, HSA and MFDFA provide a better estimation of 
 singularity spectra than SF and WL.  However, none of these methods recover the whole spectrum. We finally 
 applied all methods to the passive scalar (temperature) data set with strong ramp-cliff structure, which is an
 important signature of passive scalar turbulence \cite{Warhaft2000,Shraiman2000}.  We found that except for 
 HSA, all the methods require high-order harmonics to represent the ramp-cliff  structures. Therefore, the 
 singularity spectra provided by DFA and WL are contaminated by this large nonlinear structure. In fact, it 
 already has  been reported by several authors that for passive scalar turbulence the second-order SF and 
 Fourier power spectrum are not consistent with each other \cite{Antonia1984,Ruiz-Chavarria1996,Warhaft2000}.
\citet{Warhaft2000} stated that \lq\lq{}the reason for this is unclear, but apparently stems from the Fourier 
transform itself.''  There  is no   mathematical transform involved in SF analysis.  Now,  it seems  quite 
clear that  not only Fourier-based methods  are  strongly influenced by the ramp-cliff structure, but also SF 
analysis \cite{Huang2010PRE} and DFA.

Our experience is that the HSA is a direct measurement of scale-invariant property in the amplitude-frequency 
space. It thus requires a much larger sample size to get a convergence  result than SFs, DFA  and WL.  Thus for
  a sample of small size  without large-scale structures, SF analysis, DFA  or WL are useful to extract scaling 
 exponents and singularity spectrum since all methodologies provide almost the same result.  In this situation, the HSA is 
 useful to  catch the scaling trend. However, if the data set  possesses  a large-scale structure  (e.g., the 
 ramp-cliff structure in passive scalar turbulence, the seasonal cycle in the daily discharge of river flow, etc.),  
 we argue that   the HSA is the best choice.
 
 Finally, we would like to provide some comments  on the moment-based methods, for example, the methods presented in this work, and the singularity-based approaches  (e.g.  GMWP \cite{Turiel2006JCP,Pont2006PRE}).   Due to the convergence problem, the moment-based methods seem to require a much larger sample size than the singularity-based approaches. Thus the statistical error bars presented in this paper could be non-significant due to the finite sample size.   Another issue is the right part of the singularity spectrum, which corresponds to evaluating the negative order statistical moments for the moment-based methods.  As we already mentioned previously,  it may be inaccessible for most of the moment-based methods \cite{Turiel2006JCP}. However, the singularity-based methods can overcome this problem \cite{Turiel2006JCP,Pont2006PRE}. It seems that the idea of the singularity-based approaches can be extended into the Hilbert frame. This will be presented in a future work.

\section*{Acknowledgments}
This work is supported in part  by the National
Natural Science Foundation of China (Grant No.10772110 and No. 11072139) and the Shanghai Program for Innovative Research Team in Universities. Y.~H. was
financed in part by a  grant from the French Ministry of
Foreign Affairs and  in  part  by Universit\'e Lille. Y.H. also
acknowledges
a  financial support from EHL of Universit\'e Libre
de Bruxelles during the  preparation of  this manuscript.  We  thank  Professor P. Abry from Laboratoire de 
Physique, CNRS and ENS
Lyon (France) for providing  his wavelet leader codes. The EMD Matlab codes used in this paper are
written by Dr. Gabriel Rilling and Professor Patrick Flandrin from Laboratoire de Physique, CNRS  and ENS
Lyon (France): \footnote{http://perso.ens-lyon.fr/patrick.flandrin/emd.html}.

%\bibliographystyle{aipnum4-1}
%\bibliography{all}% Produces the bibliography via BibTeX.

\begin{thebibliography}{101}%
\makeatletter
\providecommand \@ifxundefined [1]{%
 \@ifx{#1\undefined}
}%
\providecommand \@ifnum [1]{%
 \ifnum #1\expandafter \@firstoftwo
 \else \expandafter \@secondoftwo
 \fi
}%
\providecommand \@ifx [1]{%
 \ifx #1\expandafter \@firstoftwo
 \else \expandafter \@secondoftwo
 \fi
}%
\providecommand \natexlab [1]{#1}%
\providecommand \enquote  [1]{``#1''}%
\providecommand \bibnamefont  [1]{#1}%
\providecommand \bibfnamefont [1]{#1}%
\providecommand \citenamefont [1]{#1}%
\providecommand \href@noop [0]{\@secondoftwo}%
\providecommand \href [0]{\begingroup \@sanitize@url \@href}%
\providecommand \@href[1]{\@@startlink{#1}\@@href}%
\providecommand \@@href[1]{\endgroup#1\@@endlink}%
\providecommand \@sanitize@url [0]{\catcode `\\12\catcode `\$12\catcode
  `\&12\catcode `\#12\catcode `\^12\catcode `\_12\catcode `\%12\relax}%
\providecommand \@@startlink[1]{}%
\providecommand \@@endlink[0]{}%
\providecommand \url  [0]{\begingroup\@sanitize@url \@url }%
\providecommand \@url [1]{\endgroup\@href {#1}{\urlprefix }}%
\providecommand \urlprefix  [0]{URL }%
\providecommand \Eprint [0]{\href }%
\providecommand \doibase [0]{http://dx.doi.org/}%
\providecommand \selectlanguage [0]{\@gobble}%
\providecommand \bibinfo  [0]{\@secondoftwo}%
\providecommand \bibfield  [0]{\@secondoftwo}%
\providecommand \translation [1]{[#1]}%
\providecommand \BibitemOpen [0]{}%
\providecommand \bibitemStop [0]{}%
\providecommand \bibitemNoStop [0]{.\EOS\space}%
\providecommand \EOS [0]{\spacefactor3000\relax}%
\providecommand \BibitemShut  [1]{\csname bibitem#1\endcsname}%
\let\auto@bib@innerbib\@empty
%</preamble>
\bibitem [{\citenamefont {Anselmet}\ \emph {et~al.}(1984)\citenamefont
  {Anselmet}, \citenamefont {Gagne}, \citenamefont {Hopfinger},\ and\
  \citenamefont {Antonia}}]{Anselmet1984}%
  \BibitemOpen
  \bibfield  {author} {\bibinfo {author} {\bibfnamefont {F.}~\bibnamefont
  {Anselmet}}, \bibinfo {author} {\bibfnamefont {Y.}~\bibnamefont {Gagne}},
  \bibinfo {author} {\bibfnamefont {E.~J.}\ \bibnamefont {Hopfinger}}, \ and\
  \bibinfo {author} {\bibfnamefont {R.~A.}\ \bibnamefont {Antonia}},\
  }\href@noop {} {\bibfield  {journal} {\bibinfo  {journal} {J. Fluid Mech.}\
  }\textbf {\bibinfo {volume} {140}},\ \bibinfo {pages} {63} (\bibinfo {year}
  {1984})}\BibitemShut {NoStop}%
\bibitem [{\citenamefont {Frisch}(1995)}]{Frisch1995}%
  \BibitemOpen
  \bibfield  {author} {\bibinfo {author} {\bibfnamefont {U.}~\bibnamefont
  {Frisch}},\ }\href@noop {} { {\bibinfo {title} {{Turbulence: the legacy
  of AN Kolmogorov}}}}\ (\bibinfo  {publisher} {Cambridge University Press},\
  \bibinfo {year} {1995})\BibitemShut {NoStop}%
\bibitem [{\citenamefont {Lohse}\ and\ \citenamefont {Xia}(2010)}]{Lohse2010}%
  \BibitemOpen
  \bibfield  {author} {\bibinfo {author} {\bibfnamefont {D.}~\bibnamefont
  {Lohse}}\ and\ \bibinfo {author} {\bibfnamefont {K.-Q.}\ \bibnamefont
  {Xia}},\ }\href@noop {} {\bibfield  {journal} {\bibinfo  {journal} {Ann. Rev.
  Fluid Mech.}\ }\textbf {\bibinfo {volume} {42}},\ \bibinfo {pages} {335}
  (\bibinfo {year} {2010})}\BibitemShut {NoStop}%
\bibitem [{\citenamefont {Schertzer}\ and\ \citenamefont
  {Lovejoy}(1987)}]{Schertzer1987}%
  \BibitemOpen
  \bibfield  {author} {\bibinfo {author} {\bibfnamefont {D.}~\bibnamefont
  {Schertzer}}\ and\ \bibinfo {author} {\bibfnamefont {S.}~\bibnamefont
  {Lovejoy}},\ }\href@noop {} {\bibfield  {journal} {\bibinfo  {journal} {J.
  Geophys. Res}\ }\textbf {\bibinfo {volume} {92}},\ \bibinfo {pages} {9693}
  (\bibinfo {year} {1987})}\BibitemShut {NoStop}%
\bibitem [{\citenamefont {Schmitt}\ \emph {et~al.}(1998)\citenamefont
  {Schmitt}, \citenamefont {Vannitsem},\ and\ \citenamefont
  {Barbosa}}]{Schmitt1998}%
  \BibitemOpen
  \bibfield  {author} {\bibinfo {author} {\bibfnamefont {F.~G.}\ \bibnamefont
  {Schmitt}}, \bibinfo {author} {\bibfnamefont {S.}~\bibnamefont {Vannitsem}},
  \ and\ \bibinfo {author} {\bibfnamefont {A.}~\bibnamefont {Barbosa}},\
  }\href@noop {} {\bibfield  {journal} {\bibinfo  {journal} {J. Geophys. Res.}\
  }\textbf {\bibinfo {volume} {103}},\ \bibinfo {pages} {23181} (\bibinfo
  {year} {1998})}\BibitemShut {NoStop}%
\bibitem [{\citenamefont {De~Lima}\ and\ \citenamefont
  {Grasman}(1999)}]{Delima1999}%
  \BibitemOpen
  \bibfield  {author} {\bibinfo {author} {\bibfnamefont {M.}~\bibnamefont
  {De~Lima}}\ and\ \bibinfo {author} {\bibfnamefont {J.}~\bibnamefont
  {Grasman}},\ }\href@noop {} {\bibfield  {journal} {\bibinfo  {journal} {J.
  Hydrol.}\ }\textbf {\bibinfo {volume} {220}},\ \bibinfo {pages} {1} (\bibinfo
  {year} {1999})}\BibitemShut {NoStop}%
\bibitem [{\citenamefont {Venugopal}\ \emph {et~al.}(2006)\citenamefont
  {Venugopal}, \citenamefont {Roux}, \citenamefont {Foufoula-Georgiou},\ and\
  \citenamefont {Arn{\'e}odo}}]{venugopal2006}%
  \BibitemOpen
  \bibfield  {author} {\bibinfo {author} {\bibfnamefont {V.}~\bibnamefont
  {Venugopal}}, \bibinfo {author} {\bibfnamefont {S.~G.}\ \bibnamefont {Roux}},
  \bibinfo {author} {\bibfnamefont {E.}~\bibnamefont {Foufoula-Georgiou}}, \
  and\ \bibinfo {author} {\bibfnamefont {A.}~\bibnamefont {Arn{\'e}odo}},\
  }\href@noop {} {\bibfield  {journal} {\bibinfo  {journal} {Phys. Lett. A}\
  }\textbf {\bibinfo {volume} {348}},\ \bibinfo {pages} {335} (\bibinfo {year}
  {2006})}\BibitemShut {NoStop}%
\bibitem [{\citenamefont {Ghashghaie}\ and\ \citenamefont
  {Dodge}(1996)}]{Ghashghaie1996}%
  \BibitemOpen
  \bibfield  {author} {\bibinfo {author} {\bibfnamefont {S.}~\bibnamefont
  {Ghashghaie}}\ and\ \bibinfo {author} {\bibfnamefont {Y.}~\bibnamefont
  {Dodge}},\ }\href@noop {} {\bibfield  {journal} {\bibinfo  {journal}
  {Nature}\ }\textbf {\bibinfo {volume} {381}},\ \bibinfo {pages} {27}
  (\bibinfo {year} {1996})}\BibitemShut {NoStop}%
\bibitem [{\citenamefont {Schmitt}\ \emph {et~al.}(1999)\citenamefont
  {Schmitt}, \citenamefont {Schertzer},\ and\ \citenamefont
  {Lovejoy}}]{Schmitt1999}%
  \BibitemOpen
  \bibfield  {author} {\bibinfo {author} {\bibfnamefont {F.~G.}\ \bibnamefont
  {Schmitt}}, \bibinfo {author} {\bibfnamefont {D.}~\bibnamefont {Schertzer}},
  \ and\ \bibinfo {author} {\bibfnamefont {S.}~\bibnamefont {Lovejoy}},\
  }\href@noop {} {\bibfield  {journal} {\bibinfo  {journal} {Appl. Stoch.
  Models and Data Anal.}\ }\textbf {\bibinfo {volume} {15}},\ \bibinfo {pages}
  {29} (\bibinfo {year} {1999})}\BibitemShut {NoStop}%
\bibitem [{\citenamefont {Lux}(2001)}]{Lux2001}%
  \BibitemOpen
  \bibfield  {author} {\bibinfo {author} {\bibfnamefont {T.}~\bibnamefont
  {Lux}},\ }\href@noop {} {\bibfield  {journal} {\bibinfo  {journal}
  {Quantitative Finance}\ }\textbf {\bibinfo {volume} {1}},\ \bibinfo {pages}
  {560} (\bibinfo {year} {2001})}\BibitemShut {NoStop}%
\bibitem [{\citenamefont {Calvet}\ and\ \citenamefont
  {Fisher}(2002)}]{Calvet2002}%
  \BibitemOpen
  \bibfield  {author} {\bibinfo {author} {\bibfnamefont {L.}~\bibnamefont
  {Calvet}}\ and\ \bibinfo {author} {\bibfnamefont {A.}~\bibnamefont
  {Fisher}},\ }\href@noop {} {\bibfield  {journal} {\bibinfo  {journal} {Review
  of Economics and Statistics}\ }\textbf {\bibinfo {volume} {84}},\ \bibinfo
  {pages} {381} (\bibinfo {year} {2002})}\BibitemShut {NoStop}%
\bibitem [{\citenamefont {Ivanov}\ \emph {et~al.}(1999)\citenamefont {Ivanov},
  \citenamefont {Bunde}, \citenamefont {Amaral}, \citenamefont {Havlin},
  \citenamefont {Fritsch-Yelle}, \citenamefont {Baevsky}, \citenamefont
  {Stanley},\ and\ \citenamefont {Goldberger}}]{Ivanov1999}%
  \BibitemOpen
  \bibfield  {author} {\bibinfo {author} {\bibfnamefont {P.}~\bibnamefont
  {Ivanov}}, \bibinfo {author} {\bibfnamefont {A.}~\bibnamefont {Bunde}},
  \bibinfo {author} {\bibfnamefont {L.}~\bibnamefont {Amaral}}, \bibinfo
  {author} {\bibfnamefont {S.}~\bibnamefont {Havlin}}, \bibinfo {author}
  {\bibfnamefont {J.}~\bibnamefont {Fritsch-Yelle}}, \bibinfo {author}
  {\bibfnamefont {R.}~\bibnamefont {Baevsky}}, \bibinfo {author} {\bibfnamefont
  {H.}~\bibnamefont {Stanley}}, \ and\ \bibinfo {author} {\bibfnamefont
  {A.}~\bibnamefont {Goldberger}},\ }\href@noop {} {\bibfield  {journal}
  {\bibinfo  {journal} {Europhys. Lett.}\ }\textbf {\bibinfo {volume} {48}},\
  \bibinfo {pages} {594} (\bibinfo {year} {1999})}\BibitemShut {NoStop}%
\bibitem [{\citenamefont {Monin}\ and\ \citenamefont
  {Yaglom}(1971)}]{Monin1971}%
  \BibitemOpen
  \bibfield  {author} {\bibinfo {author} {\bibfnamefont {A.~S.}\ \bibnamefont
  {Monin}}\ and\ \bibinfo {author} {\bibfnamefont {A.~M.}\ \bibnamefont
  {Yaglom}},\ }\href@noop {} { {\bibinfo {title} {{Statistical fluid
  mechanics vd II}}}}\ (\bibinfo  {publisher} {MIT Press Cambridge, Mass},\
  \bibinfo {year} {1971})\BibitemShut {NoStop}%
\bibitem [{\citenamefont {Peng}\ \emph {et~al.}(1994)\citenamefont {Peng},
  \citenamefont {Buldyrev}, \citenamefont {Havlin}, \citenamefont {Simons},
  \citenamefont {Stanley},\ and\ \citenamefont {Goldberger}}]{Peng1994PRE}%
  \BibitemOpen
  \bibfield  {author} {\bibinfo {author} {\bibfnamefont {C.K.}~\bibnamefont
  {Peng}}, \bibinfo {author} {\bibfnamefont {S.V.}~\bibnamefont {Buldyrev}},
  \bibinfo {author} {\bibfnamefont {S.}~\bibnamefont {Havlin}}, \bibinfo
  {author} {\bibfnamefont {M.}~\bibnamefont {Simons}}, \bibinfo {author}
  {\bibfnamefont {H.E.}~\bibnamefont {Stanley}}, \ and\ \bibinfo {author}
  {\bibfnamefont {A.L.}~\bibnamefont {Goldberger}},\ }\href@noop {} {\bibfield
  {journal} {\bibinfo  {journal} {Phys. Rev. E}\ }\textbf {\bibinfo {volume}
  {49}},\ \bibinfo {pages} {1685} (\bibinfo {year} {1994})}\BibitemShut
  {NoStop}%
\bibitem [{\citenamefont {Hu}\ \emph {et~al.}(2001)\citenamefont {Hu},
  \citenamefont {Ivanov}, \citenamefont {Chen}, \citenamefont {Carpena},\ and\
  \citenamefont {Eugene~Stanley}}]{Hu2001}%
  \BibitemOpen
  \bibfield  {author} {\bibinfo {author} {\bibfnamefont {K.}~\bibnamefont
  {Hu}}, \bibinfo {author} {\bibfnamefont {P.C.}~\bibnamefont {Ivanov}}, \bibinfo
  {author} {\bibfnamefont {Z.}~\bibnamefont {Chen}}, \bibinfo {author}
  {\bibfnamefont {P.}~\bibnamefont {Carpena}}, \ and\ \bibinfo {author}
  {\bibfnamefont {H.E.}~\bibnamefont {Eugene~Stanley}},\ }\href@noop {}
  {\bibfield  {journal} {\bibinfo  {journal} {Phys. Rev. E}\ }\textbf {\bibinfo
  {volume} {64}},\ \bibinfo {pages} {11114} (\bibinfo {year}
  {2001})}\BibitemShut {NoStop}%
\bibitem [{\citenamefont {Bashan}\ \emph {et~al.}(2008)\citenamefont {Bashan},
  \citenamefont {Bartsch}, \citenamefont {Kantelhardt},\ and\ \citenamefont
  {Havlin}}]{Bashan2008}%
  \BibitemOpen
  \bibfield  {author} {\bibinfo {author} {\bibfnamefont {A.}~\bibnamefont
  {Bashan}}, \bibinfo {author} {\bibfnamefont {R.}~\bibnamefont {Bartsch}},
  \bibinfo {author} {\bibfnamefont {J.}~\bibnamefont {Kantelhardt}}, \ and\
  \bibinfo {author} {\bibfnamefont {S.}~\bibnamefont {Havlin}},\ }\href@noop {}
  {\bibfield  {journal} {\bibinfo  {journal} {Physica A}\ }\textbf {\bibinfo
  {volume} {387}},\ \bibinfo {pages} {5080} (\bibinfo {year}
  {2008})}\BibitemShut {NoStop}%
\bibitem [{\citenamefont {Kantelhardt}\ \emph {et~al.}(2002)\citenamefont
  {Kantelhardt}, \citenamefont {Zschiegner}, \citenamefont {Koscielny-Bunde},
  \citenamefont {Havlin}, \citenamefont {Bunde},\ and\ \citenamefont
  {Stanley}}]{Kantelhardt2002a}%
  \BibitemOpen
  \bibfield  {author} {\bibinfo {author} {\bibfnamefont {J.}~\bibnamefont
  {Kantelhardt}}, \bibinfo {author} {\bibfnamefont {S.}~\bibnamefont
  {Zschiegner}}, \bibinfo {author} {\bibfnamefont {E.}~\bibnamefont
  {Koscielny-Bunde}}, \bibinfo {author} {\bibfnamefont {S.}~\bibnamefont
  {Havlin}}, \bibinfo {author} {\bibfnamefont {A.}~\bibnamefont {Bunde}}, \
  and\ \bibinfo {author} {\bibfnamefont {H.}~\bibnamefont {Stanley}},\
  }\href@noop {} {\bibfield  {journal} {\bibinfo  {journal} {Physica A}\
  }\textbf {\bibinfo {volume} {316}},\ \bibinfo {pages} {87} (\bibinfo {year}
  {2002})}\BibitemShut {NoStop}%
\bibitem [{\citenamefont {Muzy}\ \emph {et~al.}(1991)\citenamefont {Muzy},
  \citenamefont {Bacry},\ and\ \citenamefont {Arneodo}}]{Muzy1991PRL}%
  \BibitemOpen
  \bibfield  {author} {\bibinfo {author} {\bibfnamefont {J.F.}~\bibnamefont
  {Muzy}}, \bibinfo {author} {\bibfnamefont {E.}~\bibnamefont {Bacry}}, \ and\
  \bibinfo {author} {\bibfnamefont {A.}~\bibnamefont {Arneodo}},\ }\href@noop
  {} {\bibfield  {journal} {\bibinfo  {journal} {Phys. Rev. Lett.}\ }\textbf
  {\bibinfo {volume} {67}},\ \bibinfo {pages} {3515} (\bibinfo {year}
  {1991})}\BibitemShut {NoStop}%
\bibitem [{\citenamefont {Muzy}\ \emph {et~al.}(1993)\citenamefont {Muzy},
  \citenamefont {Bacry},\ and\ \citenamefont {Arneodo}}]{Muzy1993PRE}%
  \BibitemOpen
  \bibfield  {author} {\bibinfo {author} {\bibfnamefont {J.F.}~\bibnamefont
  {Muzy}}, \bibinfo {author} {\bibfnamefont {E.}~\bibnamefont {Bacry}}, \ and\
  \bibinfo {author} {\bibfnamefont {A.}~\bibnamefont {Arneodo}},\ }\href@noop
  {} {\bibfield  {journal} {\bibinfo  {journal} {Phys. Rev. E}\ }\textbf
  {\bibinfo {volume} {47}},\ \bibinfo {pages} {875} (\bibinfo {year}
  {1993})}\BibitemShut {NoStop}%
\bibitem [{\citenamefont {Arrault}\ \emph {et~al.}(1997)\citenamefont
  {Arrault}, \citenamefont {Arneodo}, \citenamefont {Davis},\ and\
  \citenamefont {Marshak}}]{Arrault1997PRL}%
  \BibitemOpen
  \bibfield  {author} {\bibinfo {author} {\bibfnamefont {J.}~\bibnamefont
  {Arrault}}, \bibinfo {author} {\bibfnamefont {A.}~\bibnamefont {Arneodo}},
  \bibinfo {author} {\bibfnamefont {A.}~\bibnamefont {Davis}}, \ and\ \bibinfo
  {author} {\bibfnamefont {A.}~\bibnamefont {Marshak}},\ }\href@noop {}
  {\bibfield  {journal} {\bibinfo  {journal} {Phys. Rev. Lett.}\ }\textbf
  {\bibinfo {volume} {79}},\ \bibinfo {pages} {75} (\bibinfo {year}
  {1997})}\BibitemShut {NoStop}%
\bibitem [{\citenamefont {Rodrigues~Neto}\ \emph {et~al.}(2001)\citenamefont
  {Rodrigues~Neto}, \citenamefont {Zanandrea}, \citenamefont {Ramos},
  \citenamefont {Rosa}, \citenamefont {Bolzan},\ and\ \citenamefont
  {S{\'a}}}]{Rodrigues2001}%
  \BibitemOpen
  \bibfield  {author} {\bibinfo {author} {\bibfnamefont {C.}~\bibnamefont
  {Rodrigues~Neto}}, \bibinfo {author} {\bibfnamefont {A.}~\bibnamefont
  {Zanandrea}}, \bibinfo {author} {\bibfnamefont {F.}~\bibnamefont {Ramos}},
  \bibinfo {author} {\bibfnamefont {R.}~\bibnamefont {Rosa}}, \bibinfo {author}
  {\bibfnamefont {M.}~\bibnamefont {Bolzan}}, \ and\ \bibinfo {author}
  {\bibfnamefont {L.}~\bibnamefont {S{\'a}}},\ }\href@noop {} {\bibfield
  {journal} {\bibinfo  {journal} {Physica A}\ }\textbf {\bibinfo {volume}
  {295}},\ \bibinfo {pages} {215} (\bibinfo {year} {2001})}\BibitemShut
  {NoStop}%
\bibitem [{\citenamefont {Farge}\ \emph {et~al.}(1996)\citenamefont {Farge},
  \citenamefont {Kevlahan}, \citenamefont {Perrier},\ and\ \citenamefont
  {Goirand}}]{Farge1996IEEE}%
  \BibitemOpen
  \bibfield  {author} {\bibinfo {author} {\bibfnamefont {M.}~\bibnamefont
  {Farge}}, \bibinfo {author} {\bibfnamefont {N.}~\bibnamefont {Kevlahan}},
  \bibinfo {author} {\bibfnamefont {V.}~\bibnamefont {Perrier}}, \ and\
  \bibinfo {author} {\bibfnamefont {E.}~\bibnamefont {Goirand}},\ }\href@noop
  {} {\bibfield  {journal} {\bibinfo  {journal} {IEEE J PROC}\ }\textbf
  {\bibinfo {volume} {84}},\ \bibinfo {pages} {639} (\bibinfo {year}
  {1996})}\BibitemShut {NoStop}%
\bibitem [{\citenamefont {Farge}(1992)}]{Farge1992}%
  \BibitemOpen
  \bibfield  {author} {\bibinfo {author} {\bibfnamefont {M.}~\bibnamefont
  {Farge}},\ }\href@noop {} {\bibfield  {journal} {\bibinfo  {journal} {Annu.
  Rev. Fluid Mech.}\ }\textbf {\bibinfo {volume} {24}},\ \bibinfo {pages} {395}
  (\bibinfo {year} {1992})}\BibitemShut {NoStop}%
\bibitem [{\citenamefont {Ghez}\ and\ \citenamefont
  {Vaienti}(1989)}]{Ghez1989JSP}%
  \BibitemOpen
  \bibfield  {author} {\bibinfo {author} {\bibfnamefont {J.}~\bibnamefont
  {Ghez}}\ and\ \bibinfo {author} {\bibfnamefont {S.}~\bibnamefont {Vaienti}},\
  }\href@noop {} {\bibfield  {journal} {\bibinfo  {journal} {J. Statist.
  Phys.}\ }\textbf {\bibinfo {volume} {57}},\ \bibinfo {pages} {415} (\bibinfo
  {year} {1989})}\BibitemShut {NoStop}%
\bibitem [{\citenamefont {Jaffard}\ \emph {et~al.}(2005)\citenamefont
  {Jaffard}, \citenamefont {Lashermes},\ and\ \citenamefont
  {Abry}}]{Jaffard2005wavelet}%
  \BibitemOpen
  \bibfield  {author} {\bibinfo {author} {\bibfnamefont {S.}~\bibnamefont
  {Jaffard}}, \bibinfo {author} {\bibfnamefont {B.}~\bibnamefont {Lashermes}},
  \ and\ \bibinfo {author} {\bibfnamefont {P.}~\bibnamefont {Abry}},\
  }\href@noop {} {\bibfield  {journal} {\bibinfo  {journal} {Wavelet Analysis
  and Applications}\ } (\bibinfo {year} {2005})}\BibitemShut {NoStop}%
\bibitem [{\citenamefont {Lashermes}\ \emph {et~al.}(2008)\citenamefont
  {Lashermes}, \citenamefont {Roux}, \citenamefont {Abry},\ and\ \citenamefont
  {Jaffard}}]{Lashermes2008EPJB}%
  \BibitemOpen
  \bibfield  {author} {\bibinfo {author} {\bibfnamefont {B.}~\bibnamefont
  {Lashermes}}, \bibinfo {author} {\bibfnamefont {S.}~\bibnamefont {Roux}},
  \bibinfo {author} {\bibfnamefont {P.}~\bibnamefont {Abry}}, \ and\ \bibinfo
  {author} {\bibfnamefont {S.}~\bibnamefont {Jaffard}},\ }\href@noop {}
  {\bibfield  {journal} {\bibinfo  {journal} {EPJB}\
  }\textbf {\bibinfo {volume} {61}},\ \bibinfo {pages} {201} (\bibinfo {year}
  {2008})}\BibitemShut {NoStop}%
\bibitem [{\citenamefont {Serrano}\ and\ \citenamefont
  {Figliola}(2009)}]{Serrano2009}%
  \BibitemOpen
  \bibfield  {author} {\bibinfo {author} {\bibfnamefont {E.}~\bibnamefont
  {Serrano}}\ and\ \bibinfo {author} {\bibfnamefont {A.}~\bibnamefont
  {Figliola}},\ }\href@noop {} {\bibfield  {journal} {\bibinfo  {journal}
  {Physica A}\ }\textbf {\bibinfo {volume} {388}},\ \bibinfo {pages} {2793}
  (\bibinfo {year} {2009})}\BibitemShut {NoStop}%
\bibitem [{\citenamefont {Lashermes}\ \emph {et~al.}(2005)\citenamefont
  {Lashermes}, \citenamefont {Jaffard},\ and\ \citenamefont
  {Abry}}]{Lashermes2005wavelet}%
  \BibitemOpen
  \bibfield  {author} {\bibinfo {author} {\bibfnamefont {B.}~\bibnamefont
  {Lashermes}}, \bibinfo {author} {\bibfnamefont {S.}~\bibnamefont {Jaffard}},
  \ and\ \bibinfo {author} {\bibfnamefont {P.}~\bibnamefont {Abry}},\
  }\bibfield  {booktitle} { {\bibinfo {booktitle} {ICASSP 2005,
  Philadelphia, USA}},\ }\href@noop {} {\  (\bibinfo {year}
  {2005})}\BibitemShut {NoStop}%
\bibitem [{\citenamefont {Pont}\ \emph {et~al.}(2006)\citenamefont {Pont},
  \citenamefont {Turiel},\ and\ \citenamefont
  {P{\'e}rez-Vicente}}]{Pont2006PRE}%
  \BibitemOpen
  \bibfield  {author} {\bibinfo {author} {\bibfnamefont {O.}~\bibnamefont
  {Pont}}, \bibinfo {author} {\bibfnamefont {A.}~\bibnamefont {Turiel}}, \ and\
  \bibinfo {author} {\bibfnamefont {C.}~\bibnamefont {P{\'e}rez-Vicente}},\
  }\href@noop {} {\bibfield  {journal} {\bibinfo  {journal} {Physical Review
  E}\ }\textbf {\bibinfo {volume} {74}},\ \bibinfo {pages} {061110} (\bibinfo
  {year} {2006})}\BibitemShut {NoStop}%
\bibitem [{\citenamefont {Turiel}\ \emph {et~al.}(2006)\citenamefont {Turiel},
  \citenamefont {P{\'e}rez-Vicente},\ and\ \citenamefont
  {Grazzini}}]{Turiel2006JCP}%
  \BibitemOpen
  \bibfield  {author} {\bibinfo {author} {\bibfnamefont {A.}~\bibnamefont
  {Turiel}}, \bibinfo {author} {\bibfnamefont {C.J.}~\bibnamefont
  {P{\'e}rez-Vicente}}, \ and\ \bibinfo {author} {\bibfnamefont
  {J.}~\bibnamefont {Grazzini}},\ }\href@noop {} {\bibfield  {journal}
  {\bibinfo  {journal} {J. Comput. Phys.}\ }\textbf {\bibinfo
  {volume} {216}},\ \bibinfo {pages} {362} (\bibinfo {year}
  {2006})}\BibitemShut {NoStop}%
\bibitem [{\citenamefont {Huang}\ \emph
  {et~al.}(2008{\natexlab{a}})\citenamefont {Huang}, \citenamefont {Schmitt},
  \citenamefont {Lu},\ and\ \citenamefont {Liu}}]{Huang2008EPL}%
  \BibitemOpen
  \bibfield  {author} {\bibinfo {author} {\bibfnamefont {Y.}~\bibnamefont
  {Huang}}, \bibinfo {author} {\bibfnamefont {F.~G.}\ \bibnamefont {Schmitt}},
  \bibinfo {author} {\bibfnamefont {Z.}~\bibnamefont {Lu}}, \ and\ \bibinfo
  {author} {\bibfnamefont {Y.}~\bibnamefont {Liu}},\ }\href@noop {} {\bibfield
  {journal} {\bibinfo  {journal} {Europhys. Lett.}\ }\textbf {\bibinfo {volume}
  {84}},\ \bibinfo {pages} {40010} (\bibinfo {year}
  {2008}{\natexlab{a}})}\BibitemShut {NoStop}%
\bibitem [{\citenamefont {Huang}\ \emph
  {et~al.}(2008{\natexlab{b}})\citenamefont {Huang}, \citenamefont {Schmitt},
  \citenamefont {Lu},\ and\ \citenamefont {Liu}}]{Huang2008TSI}%
  \BibitemOpen
  \bibfield  {author} {\bibinfo {author} {\bibfnamefont {Y.}~\bibnamefont
  {Huang}}, \bibinfo {author} {\bibfnamefont {F.~G.}\ \bibnamefont {Schmitt}},
  \bibinfo {author} {\bibfnamefont {Z.}~\bibnamefont {Lu}}, \ and\ \bibinfo
  {author} {\bibfnamefont {Y.}~\bibnamefont {Liu}},\ }\href@noop {} {\bibfield
  {journal} {\bibinfo  {journal} {Traitement du Signal}\ }\textbf {\bibinfo
  {volume} {25}},\ \bibinfo {pages} {481} (\bibinfo {year}
  {2008}{\natexlab{b}})}\BibitemShut {NoStop}%
\bibitem [{\citenamefont {Huang}(2009)}]{Huang2009PHD}%
  \BibitemOpen
  \bibfield  {author} {\bibinfo {author} {\bibfnamefont {Y.}~\bibnamefont
  {Huang}},\ }  {\bibinfo {title} {Arbitrary Order Hilbert Spectral
  Analysis: Definition and Application to fully developed turbulence and
  environmental time series}},\ \href@noop {} {Ph.D. thesis},\ \bibinfo
  {school} {Universit\'e des Sciences et Technologies de Lille - Lille 1,
  France \& Shanghai University, China} (\bibinfo {year} {2009});{http://tel.archives-ouvertes.fr/tel-00439605/fr}.\BibitemShut
  {NoStop}%
\bibitem [{\citenamefont {Schmitt}\ \emph {et~al.}(2009)\citenamefont
  {Schmitt}, \citenamefont {Huang}, \citenamefont {Lu}, \citenamefont {Y.},\
  and\ \citenamefont {Fernandez}}]{Schmitt2009JMS}%
  \BibitemOpen
  \bibfield  {author} {\bibinfo {author} {\bibfnamefont {F.~G.}\ \bibnamefont
  {Schmitt}}, \bibinfo {author} {\bibfnamefont {Y.}~\bibnamefont {Huang}},
  \bibinfo {author} {\bibfnamefont {Z.}~\bibnamefont {Lu}}, \bibinfo {author}
  {\bibfnamefont {L.}~\bibnamefont {Y.}}, \ and\ \bibinfo {author}
  {\bibfnamefont {N.}~\bibnamefont {Fernandez}},\ }\href@noop {} {\bibfield
  {journal} {\bibinfo  {journal} {J. Mar. Sys.}\ }\textbf {\bibinfo {volume}
  {77}},\ \bibinfo {pages} {473} (\bibinfo {year} {2009})}\BibitemShut
  {NoStop}%
\bibitem [{\citenamefont {Huang}\ \emph {et~al.}(2009)\citenamefont {Huang},
  \citenamefont {Schmitt}, \citenamefont {Lu},\ and\ \citenamefont
  {Liu}}]{Huang2009Hydrol}%
  \BibitemOpen
  \bibfield  {author} {\bibinfo {author} {\bibfnamefont {Y.}~\bibnamefont
  {Huang}}, \bibinfo {author} {\bibfnamefont {F.~G.}\ \bibnamefont {Schmitt}},
  \bibinfo {author} {\bibfnamefont {Z.}~\bibnamefont {Lu}}, \ and\ \bibinfo
  {author} {\bibfnamefont {Y.}~\bibnamefont {Liu}},\ }\href@noop {} {\bibfield
  {journal} {\bibinfo  {journal} {J. Hydrol.}\ }\textbf {\bibinfo {volume}
  {373}},\ \bibinfo {pages} {103} (\bibinfo {year} {2009})}\BibitemShut
  {NoStop}%
\bibitem [{\citenamefont {Huang}\ \emph {et~al.}(2010)\citenamefont {Huang},
  \citenamefont {Schmitt}, \citenamefont {Lu}, \citenamefont {Fougairolles},
  \citenamefont {Gagne},\ and\ \citenamefont {Liu}}]{Huang2010PRE}%
  \BibitemOpen
  \bibfield  {author} {\bibinfo {author} {\bibfnamefont {Y.}~\bibnamefont
  {Huang}}, \bibinfo {author} {\bibfnamefont {F.}~\bibnamefont {Schmitt}},
  \bibinfo {author} {\bibfnamefont {Z.}~\bibnamefont {Lu}}, \bibinfo {author}
  {\bibfnamefont {P.}~\bibnamefont {Fougairolles}}, \bibinfo {author}
  {\bibfnamefont {Y.}~\bibnamefont {Gagne}}, \ and\ \bibinfo {author}
  {\bibfnamefont {Y.}~\bibnamefont {Liu}},\ }\href@noop {} {\bibfield
  {journal} {\bibinfo  {journal} {Phys. Rev. E}\ }\textbf {\bibinfo {volume}
  {82}},\ \bibinfo {pages} {26319} (\bibinfo {year} {2010})}\BibitemShut
  {NoStop}%
\bibitem [{\citenamefont {Cohen}(1995)}]{Cohen1995}%
  \BibitemOpen
  \bibfield  {author} {\bibinfo {author} {\bibfnamefont {L.}~\bibnamefont
  {Cohen}},\ }\href@noop {} { {\bibinfo {title} {{Time-frequency
  analysis}}}}\ (\bibinfo  {publisher} {Prentice Hall PTR Englewood Cliffs,
  NJ},\ \bibinfo {year} {1995})\BibitemShut {NoStop}%
\bibitem [{\citenamefont {Huang}\ \emph {et~al.}(1998)\citenamefont {Huang},
  \citenamefont {Shen}, \citenamefont {Long}, \citenamefont {Wu}, \citenamefont
  {Shih}, \citenamefont {Zheng}, \citenamefont {Yen}, \citenamefont {Tung},\
  and\ \citenamefont {Liu}}]{Huang1998EMD}%
  \BibitemOpen
  \bibfield  {author} {\bibinfo {author} {\bibfnamefont {N.~E.}\ \bibnamefont
  {Huang}}, \bibinfo {author} {\bibfnamefont {Z.}~\bibnamefont {Shen}},
  \bibinfo {author} {\bibfnamefont {S.~R.}\ \bibnamefont {Long}}, \bibinfo
  {author} {\bibfnamefont {M.~C.}\ \bibnamefont {Wu}}, \bibinfo {author}
  {\bibfnamefont {H.~H.}\ \bibnamefont {Shih}}, \bibinfo {author}
  {\bibfnamefont {Q.}~\bibnamefont {Zheng}}, \bibinfo {author} {\bibfnamefont
  {N.}~\bibnamefont {Yen}}, \bibinfo {author} {\bibfnamefont {C.~C.}\
  \bibnamefont {Tung}}, \ and\ \bibinfo {author} {\bibfnamefont {H.~H.}\
  \bibnamefont {Liu}},\ }\href@noop {} {\bibfield  {journal} {\bibinfo
  {journal} {Proc. R. Soc. London, Ser. A}\ }\textbf {\bibinfo {volume}
  {454}},\ \bibinfo {pages} {903} (\bibinfo {year} {1998})}\BibitemShut
  {NoStop}%
\bibitem [{\citenamefont {Huang}\ \emph {et~al.}(1999)\citenamefont {Huang},
  \citenamefont {Shen},\ and\ \citenamefont {Long}}]{Huang1999EMD}%
  \BibitemOpen
  \bibfield  {author} {\bibinfo {author} {\bibfnamefont {N.~E.}\ \bibnamefont
  {Huang}}, \bibinfo {author} {\bibfnamefont {Z.}~\bibnamefont {Shen}}, \ and\
  \bibinfo {author} {\bibfnamefont {S.~R.}\ \bibnamefont {Long}},\ }\href@noop
  {} {\bibfield  {journal} {\bibinfo  {journal} {Annu. Rev. Fluid Mech.}\
  }\textbf {\bibinfo {volume} {31}},\ \bibinfo {pages} {417} (\bibinfo {year}
  {1999})}\BibitemShut {NoStop}%
\bibitem [{\citenamefont {Rilling}\ \emph {et~al.}(2003)\citenamefont
  {Rilling}, \citenamefont {Flandrin},\ and\ \citenamefont
  {Gon\c{c}alv\`es}}]{Rilling2003EMD}%
  \BibitemOpen
  \bibfield  {author} {\bibinfo {author} {\bibfnamefont {G.}~\bibnamefont
  {Rilling}}, \bibinfo {author} {\bibfnamefont {P.}~\bibnamefont {Flandrin}}, \
  and\ \bibinfo {author} {\bibfnamefont {P.}~\bibnamefont {Gon\c{c}alv\`es}},\
  }\href@noop {} {\bibfield  {journal} {\bibinfo  {journal} {IEEE-EURASIP
  Workshop on Nonlinear Signal and Image Processing}\ } (\bibinfo {year}
  {2003})}\BibitemShut {NoStop}%
\bibitem [{\citenamefont {Flandrin}\ and\ \citenamefont
  {Gon{\c{c}}alv\`es}(2004)}]{Flandrin2004EMDa}%
  \BibitemOpen
  \bibfield  {author} {\bibinfo {author} {\bibfnamefont {P.}~\bibnamefont
  {Flandrin}}\ and\ \bibinfo {author} {\bibfnamefont {P.}~\bibnamefont
  {Gon{\c{c}}alv\`es}},\ }\href@noop {} {\bibfield  {journal} {\bibinfo
  {journal} {Int. J. Wavelets, Multires. Info. Proc.}\ }\textbf {\bibinfo
  {volume} {2}},\ \bibinfo {pages} {477} (\bibinfo {year} {2004})}\BibitemShut
  {NoStop}%
\bibitem [{\citenamefont {Huang}\ \emph
  {et~al.}(2003{\natexlab{a}})\citenamefont {Huang}, \citenamefont {Wu},
  \citenamefont {Long}, \citenamefont {Shen}, \citenamefont {Qu}, \citenamefont
  {Gloersen},\ and\ \citenamefont {Fan}}]{Huang2003b}%
  \BibitemOpen
  \bibfield  {author} {\bibinfo {author} {\bibfnamefont {N.~E.}\ \bibnamefont
  {Huang}}, \bibinfo {author} {\bibfnamefont {M.~L.}\ \bibnamefont {Wu}},
  \bibinfo {author} {\bibfnamefont {S.~R.}\ \bibnamefont {Long}}, \bibinfo
  {author} {\bibfnamefont {S.~S.~P.}\ \bibnamefont {Shen}}, \bibinfo {author}
  {\bibfnamefont {W.}~\bibnamefont {Qu}}, \bibinfo {author} {\bibfnamefont
  {P.}~\bibnamefont {Gloersen}}, \ and\ \bibinfo {author} {\bibfnamefont
  {K.~L.}\ \bibnamefont {Fan}},\ }\href@noop {} {\bibfield  {journal} {\bibinfo
   {journal} {Proc. R. Soc. London, Ser. A}\ }\textbf {\bibinfo {volume}
  {459}},\ \bibinfo {pages} {2317} (\bibinfo {year}
  {2003}{\natexlab{a}})}\BibitemShut {NoStop}%
\bibitem [{\citenamefont {Huang}(2005)}]{Huang2005EMDa}%
  \BibitemOpen
  \bibfield  {author} {\bibinfo {author} {\bibfnamefont {N.~E.}\ \bibnamefont
  {Huang}},\ } {\bibinfo {title} {Hilbert-huang transform and its
  applications},}\ \ (\bibinfo  {publisher} {World Scientific,  Singapore},\ \bibinfo
  {year} {2005})\ Chap.\ \bibinfo {chapter} {1}, pp.\ \bibinfo
  {pages} {1--26}\BibitemShut {NoStop}%
\bibitem [{\citenamefont {Hwang}\ \emph {et~al.}(2003)\citenamefont {Hwang},
  \citenamefont {Huang},\ and\ \citenamefont {Wang}}]{Hwang2003}%
  \BibitemOpen
  \bibfield  {author} {\bibinfo {author} {\bibfnamefont {P.~A.}\ \bibnamefont
  {Hwang}}, \bibinfo {author} {\bibfnamefont {N.~E.}\ \bibnamefont {Huang}}, \
  and\ \bibinfo {author} {\bibfnamefont {D.~W.}\ \bibnamefont {Wang}},\
  }\href@noop {} {\bibfield  {journal} {\bibinfo  {journal} {Appl. Ocean Res.}\
  }\textbf {\bibinfo {volume} {25}},\ \bibinfo {pages} {187} (\bibinfo {year}
  {2003})}\BibitemShut {NoStop}%
\bibitem [{\citenamefont {Veltcheva}\ and\ \citenamefont
  {Soares}(2004)}]{Veltcheva2004}%
  \BibitemOpen
  \bibfield  {author} {\bibinfo {author} {\bibfnamefont {A.~D.}\ \bibnamefont
  {Veltcheva}}\ and\ \bibinfo {author} {\bibfnamefont {C.~G.}\ \bibnamefont
  {Soares}},\ }\href@noop {} {\bibfield  {journal} {\bibinfo  {journal} {Appl.
  Ocean Res.}\ }\textbf {\bibinfo {volume} {26}},\ \bibinfo {pages} {1}
  (\bibinfo {year} {2004})}\BibitemShut {NoStop}%
\bibitem [{\citenamefont {Echeverria}\ \emph {et~al.}(2001)\citenamefont
  {Echeverria}, \citenamefont {Crowe}, \citenamefont {Woolfson},\ and\
  \citenamefont {Hayes-Gill}}]{Echeverria2001}%
  \BibitemOpen
  \bibfield  {author} {\bibinfo {author} {\bibfnamefont {J.~C.}\ \bibnamefont
  {Echeverria}}, \bibinfo {author} {\bibfnamefont {J.~A.}\ \bibnamefont
  {Crowe}}, \bibinfo {author} {\bibfnamefont {M.~S.}\ \bibnamefont {Woolfson}},
  \ and\ \bibinfo {author} {\bibfnamefont {B.~R.}\ \bibnamefont {Hayes-Gill}},\
  }\href@noop {} {\bibfield  {journal} {\bibinfo  {journal} {Med. Biol. Eng.
  Comput.}\ }\textbf {\bibinfo {volume} {39}},\ \bibinfo {pages} {471}
  (\bibinfo {year} {2001})}\BibitemShut {NoStop}%
\bibitem [{\citenamefont {Balocchi}\ \emph {et~al.}(2004)\citenamefont
  {Balocchi}, \citenamefont {Menicucci}, \citenamefont {Santarcangelo},
  \citenamefont {Sebastiani}, \citenamefont {Gemignani}, \citenamefont
  {Ghelarducci},\ and\ \citenamefont {Varanini}}]{Balocchi2004}%
  \BibitemOpen
  \bibfield  {author} {\bibinfo {author} {\bibfnamefont {R.}~\bibnamefont
  {Balocchi}}, \bibinfo {author} {\bibfnamefont {D.}~\bibnamefont {Menicucci}},
  \bibinfo {author} {\bibfnamefont {E.}~\bibnamefont {Santarcangelo}}, \bibinfo
  {author} {\bibfnamefont {L.}~\bibnamefont {Sebastiani}}, \bibinfo {author}
  {\bibfnamefont {A.}~\bibnamefont {Gemignani}}, \bibinfo {author}
  {\bibfnamefont {B.}~\bibnamefont {Ghelarducci}}, \ and\ \bibinfo {author}
  {\bibfnamefont {M.}~\bibnamefont {Varanini}},\ }\href@noop {} {\bibfield
  {journal} {\bibinfo  {journal} {Chaos Soliton Fract.}\ }\textbf {\bibinfo
  {volume} {20}},\ \bibinfo {pages} {171} (\bibinfo {year} {2004})}\BibitemShut
  {NoStop}%
\bibitem [{\citenamefont {Ponomarenko}\ \emph {et~al.}(2005)\citenamefont
  {Ponomarenko}, \citenamefont {Prokhorov}, \citenamefont {Bespyatov},
  \citenamefont {Bodrov},\ and\ \citenamefont {Gridnev}}]{Ponomarenko2005}%
  \BibitemOpen
  \bibfield  {author} {\bibinfo {author} {\bibfnamefont {V.~I.}\ \bibnamefont
  {Ponomarenko}}, \bibinfo {author} {\bibfnamefont {M.~D.}\ \bibnamefont
  {Prokhorov}}, \bibinfo {author} {\bibfnamefont {A.~B.}\ \bibnamefont
  {Bespyatov}}, \bibinfo {author} {\bibfnamefont {M.~B.}\ \bibnamefont
  {Bodrov}}, \ and\ \bibinfo {author} {\bibfnamefont {V.~I.}\ \bibnamefont
  {Gridnev}},\ }\href@noop {} {\bibfield  {journal} {\bibinfo  {journal} {Chaos
  Soliton Fract.}\ }\textbf {\bibinfo {volume} {23}},\ \bibinfo {pages} {1429}
  (\bibinfo {year} {2005})}\BibitemShut {NoStop}%
\bibitem [{\citenamefont {Huang}\ \emph
  {et~al.}(2003{\natexlab{b}})\citenamefont {Huang}, \citenamefont {Wu},
  \citenamefont {Qu}, \citenamefont {Long},\ and\ \citenamefont
  {Shen}}]{Huang2003a}%
  \BibitemOpen
  \bibfield  {author} {\bibinfo {author} {\bibfnamefont {N.~E.}\ \bibnamefont
  {Huang}}, \bibinfo {author} {\bibfnamefont {M.~L.}\ \bibnamefont {Wu}},
  \bibinfo {author} {\bibfnamefont {W.}~\bibnamefont {Qu}}, \bibinfo {author}
  {\bibfnamefont {S.~R.}\ \bibnamefont {Long}}, \ and\ \bibinfo {author}
  {\bibfnamefont {S.~S.~P.}\ \bibnamefont {Shen}},\ }\href@noop {} {\bibfield
  {journal} {\bibinfo  {journal} {Appl. Stoch. Model Bus.}\ }\textbf {\bibinfo
  {volume} {19}},\ \bibinfo {pages} {245} (\bibinfo {year}
  {2003}{\natexlab{b}})}\BibitemShut {NoStop}%
\bibitem [{\citenamefont {Coughlin}\ and\ \citenamefont
  {Tung}(2004)}]{Coughlin2004}%
  \BibitemOpen
  \bibfield  {author} {\bibinfo {author} {\bibfnamefont {K.~T.}\ \bibnamefont
  {Coughlin}}\ and\ \bibinfo {author} {\bibfnamefont {K.~K.}\ \bibnamefont
  {Tung}},\ }\href@noop {} {\bibfield  {journal} {\bibinfo  {journal} {Adv.
  Space Res.}\ }\textbf {\bibinfo {volume} {34}},\ \bibinfo {pages} {323}
  (\bibinfo {year} {2004})}\BibitemShut {NoStop}%
\bibitem [{\citenamefont {J{\'a}nosi}\ and\ \citenamefont
  {M{\"u}ller}(2005)}]{Janosi2005}%
  \BibitemOpen
  \bibfield  {author} {\bibinfo {author} {\bibfnamefont {I.M.}~\bibnamefont
  {J{\'a}nosi}}\ and\ \bibinfo {author} {\bibfnamefont {R.}~\bibnamefont
  {M{\"u}ller}},\ }\href@noop {} {\bibfield  {journal} {\bibinfo  {journal}
  {Phys. Rev. E}\ }\textbf {\bibinfo {volume} {71}},\ \bibinfo {pages} {56126}
  (\bibinfo {year} {2005})}\BibitemShut {NoStop}%
\bibitem [{\citenamefont {Molla}\ \emph {et~al.}(2006)\citenamefont {Molla},
  \citenamefont {Rahman}, \citenamefont {Sumi},\ and\ \citenamefont
  {Banik}}]{Molla2006a}%
  \BibitemOpen
  \bibfield  {author} {\bibinfo {author} {\bibfnamefont {M.~K.~I.}\
  \bibnamefont {Molla}}, \bibinfo {author} {\bibfnamefont {M.~S.}\ \bibnamefont
  {Rahman}}, \bibinfo {author} {\bibfnamefont {A.}~\bibnamefont {Sumi}}, \ and\
  \bibinfo {author} {\bibfnamefont {P.}~\bibnamefont {Banik}},\ }\href@noop {}
  {\bibfield  {journal} {\bibinfo  {journal} {Discrete Dyn. Nat. Soc.}\
  }\textbf {\bibinfo {volume} {2006}},\ \bibinfo {pages} {Article ID 45348, 17
  pages} (\bibinfo {year} {2006})},\ \bibinfo {note}
  {doi:10.1155/DDNS/2006/45348}\BibitemShut {NoStop}%
\bibitem [{\citenamefont {Sol{\'e}}\ \emph {et~al.}(2007)\citenamefont
  {Sol{\'e}}, \citenamefont {Turiel},\ and\ \citenamefont {Llebot}}]{Sole2007}%
  \BibitemOpen
  \bibfield  {author} {\bibinfo {author} {\bibfnamefont {J.}~\bibnamefont
  {Sol{\'e}}}, \bibinfo {author} {\bibfnamefont {A.}~\bibnamefont {Turiel}}, \
  and\ \bibinfo {author} {\bibfnamefont {J.}~\bibnamefont {Llebot}},\
  }\href@noop {} {\bibfield  {journal} {\bibinfo  {journal} {Nat. Hazard Earth
  Sys. Sci.}\ }\textbf {\bibinfo {volume} {7}},\ \bibinfo {pages} {299}
  (\bibinfo {year} {2007})}\BibitemShut {NoStop}%
\bibitem [{\citenamefont {Wu}\ \emph {et~al.}(2007)\citenamefont {Wu},
  \citenamefont {Huang}, \citenamefont {Long},\ and\ \citenamefont
  {Peng}}]{Wu2007}%
  \BibitemOpen
  \bibfield  {author} {\bibinfo {author} {\bibfnamefont {Z.}~\bibnamefont
  {Wu}}, \bibinfo {author} {\bibfnamefont {N.~E.}\ \bibnamefont {Huang}},
  \bibinfo {author} {\bibfnamefont {S.~R.}\ \bibnamefont {Long}}, \ and\
  \bibinfo {author} {\bibfnamefont {C.}~\bibnamefont {Peng}},\ }\href@noop {}
  {\bibfield  {journal} {\bibinfo  {journal} {PNAS}\ }\textbf {\bibinfo
  {volume} {104}},\ \bibinfo {pages} {14889} (\bibinfo {year}
  {2007})}\BibitemShut {NoStop}%
\bibitem [{\citenamefont {Loh}\ \emph {et~al.}(2001)\citenamefont {Loh},
  \citenamefont {Wu},\ and\ \citenamefont {Huang}}]{Loh2001}%
  \BibitemOpen
  \bibfield  {author} {\bibinfo {author} {\bibfnamefont {C.~H.}\ \bibnamefont
  {Loh}}, \bibinfo {author} {\bibfnamefont {T.~C.}\ \bibnamefont {Wu}}, \ and\
  \bibinfo {author} {\bibfnamefont {N.~E.}\ \bibnamefont {Huang}},\ }\href@noop
  {} {\bibfield  {journal} {\bibinfo  {journal} {BSSA}\ }\textbf {\bibinfo
  {volume} {91}},\ \bibinfo {pages} {1339} (\bibinfo {year}
  {2001})}\BibitemShut {NoStop}%
\bibitem [{\citenamefont {Chen}\ \emph {et~al.}(2004)\citenamefont {Chen},
  \citenamefont {Xu},\ and\ \citenamefont {Zhang}}]{Chen2004}%
  \BibitemOpen
  \bibfield  {author} {\bibinfo {author} {\bibfnamefont {J.}~\bibnamefont
  {Chen}}, \bibinfo {author} {\bibfnamefont {Y.~L.}\ \bibnamefont {Xu}}, \ and\
  \bibinfo {author} {\bibfnamefont {R.~C.}\ \bibnamefont {Zhang}},\ }\href@noop
  {} {\bibfield  {journal} {\bibinfo  {journal} {J. Wind Eng. Ind. Aerodyn.}\
  }\textbf {\bibinfo {volume} {92}},\ \bibinfo {pages} {805} (\bibinfo {year}
  {2004})}\BibitemShut {NoStop}%
\bibitem [{\citenamefont {Loutridis}(2005)}]{Loutridis2005}%
  \BibitemOpen
  \bibfield  {author} {\bibinfo {author} {\bibfnamefont {S.~J.}\ \bibnamefont
  {Loutridis}},\ }\href@noop {} {\bibfield  {journal} {\bibinfo  {journal}
  {Appl. Acoust.}\ }\textbf {\bibinfo {volume} {66}},\ \bibinfo {pages} {1399}
  (\bibinfo {year} {2005})}\BibitemShut {NoStop}%
\bibitem [{\citenamefont {Schmitt}\ \emph {et~al.}(2007)\citenamefont
  {Schmitt}, \citenamefont {Huang}, \citenamefont {Lu}, \citenamefont {Zongo},
  \citenamefont {Molinero},\ and\ \citenamefont {Liu}}]{Schmitt2007}%
  \BibitemOpen
  \bibfield  {author} {\bibinfo {author} {\bibfnamefont {F.~G.}\ \bibnamefont
  {Schmitt}}, \bibinfo {author} {\bibfnamefont {Y.}~\bibnamefont {Huang}},
  \bibinfo {author} {\bibfnamefont {Z.}~\bibnamefont {Lu}}, \bibinfo {author}
  {\bibfnamefont {S.~B.}\ \bibnamefont {Zongo}}, \bibinfo {author}
  {\bibfnamefont {J.~C.}\ \bibnamefont {Molinero}}, \ and\ \bibinfo {author}
  {\bibfnamefont {Y.}~\bibnamefont {Liu}},\ }in\ \href@noop {} {{\bibinfo
  {booktitle} {Nonlinear Dynamics in Geosciences. edited by A. Tsonis and J.
  Elsner}}}\ (\bibinfo  {publisher} {Springer},\ \bibinfo {year} {2007})\ pp.\
  \bibinfo {pages} {261--280}\BibitemShut {NoStop}%
\bibitem [{\citenamefont {Flandrin}\ \emph {et~al.}(2004)\citenamefont
  {Flandrin}, \citenamefont {Rilling},\ and\ \citenamefont
  {Gon{\c{c}}alv\`es}}]{Flandrin2004EMDb}%
  \BibitemOpen
  \bibfield  {author} {\bibinfo {author} {\bibfnamefont {P.}~\bibnamefont
  {Flandrin}}, \bibinfo {author} {\bibfnamefont {G.}~\bibnamefont {Rilling}}, \
  and\ \bibinfo {author} {\bibfnamefont {P.}~\bibnamefont
  {Gon{\c{c}}alv\`es}},\ }\href@noop {} {\bibfield  {journal} {\bibinfo
  {journal} {IEEE Sig. Proc. Lett.}\ }\textbf {\bibinfo {volume} {11}},\
  \bibinfo {pages} {112} (\bibinfo {year} {2004})}\BibitemShut {NoStop}%
\bibitem [{\citenamefont {Long}\ \emph {et~al.}(1995)\citenamefont {Long},
  \citenamefont {Huang}, \citenamefont {Tung}, \citenamefont {Wu},
  \citenamefont {Lin}, \citenamefont {Mollo-Christensen},\ and\ \citenamefont
  {Yuan}}]{Long1995}%
  \BibitemOpen
  \bibfield  {author} {\bibinfo {author} {\bibfnamefont {S.~R.}\ \bibnamefont
  {Long}}, \bibinfo {author} {\bibfnamefont {N.~E.}\ \bibnamefont {Huang}},
  \bibinfo {author} {\bibfnamefont {C.~C.}\ \bibnamefont {Tung}}, \bibinfo
  {author} {\bibfnamefont {M.~L.}\ \bibnamefont {Wu}}, \bibinfo {author}
  {\bibfnamefont {R.~Q.}\ \bibnamefont {Lin}}, \bibinfo {author} {\bibfnamefont
  {E.}~\bibnamefont {Mollo-Christensen}}, \ and\ \bibinfo {author}
  {\bibfnamefont {Y.}~\bibnamefont {Yuan}},\ }\href@noop {} {\bibfield
  {journal} {\bibinfo  {journal} {IEEE Geoscience and Remote Sensing Soc.
  Lett.}\ }\textbf {\bibinfo {volume} {3}},\ \bibinfo {pages} {6} (\bibinfo
  {year} {1995})}\BibitemShut {NoStop}%
\bibitem [{\citenamefont {Flandrin}(1998)}]{Flandrin1998}%
  \BibitemOpen
  \bibfield  {author} {\bibinfo {author} {\bibfnamefont {P.}~\bibnamefont
  {Flandrin}},\ }\href@noop {} {{\bibinfo {title}
  {{Time-frequency/time-scale analysis}}}}\ (\bibinfo  {publisher} {Academic
  Press},\ \bibinfo {year} {1998})\BibitemShut {NoStop}%
\bibitem [{Note1()}]{Note1}%
  \BibitemOpen
  \bibinfo {note} {In fact, the Eq.~(\ref {eq:arbitrary}) is convergence when
  $q\ge -1$. However, in practice, we only consider the case $q\ge
  0$.}\BibitemShut {Stop}%
\bibitem [{\citenamefont {Rilling}\ and\ \citenamefont
  {Flandrin}(2006)}]{Rilling2006}%
  \BibitemOpen
  \bibfield  {author} {\bibinfo {author} {\bibfnamefont {G.}~\bibnamefont
  {Rilling}}\ and\ \bibinfo {author} {\bibfnamefont {P.}~\bibnamefont
  {Flandrin}},\ }\href@noop {} {\bibfield  {journal} {\bibinfo  {journal} {IEEE
  International Conference on Acoustics, Speech and Signal Processing, 2006.
  ICASSP 2006 Proceedings. 2006}\ }\textbf {\bibinfo {volume} {3}},\ \bibinfo
  {pages} {444} (\bibinfo {year} {2006})}\BibitemShut {NoStop}%
\bibitem [{\citenamefont {Rilling}\ and\ \citenamefont
  {Flandrin}(2008)}]{Rilling2008}%
  \BibitemOpen
  \bibfield  {author} {\bibinfo {author} {\bibfnamefont {G.}~\bibnamefont
  {Rilling}}\ and\ \bibinfo {author} {\bibfnamefont {P.}~\bibnamefont
  {Flandrin}},\ }\href@noop {} {\bibfield  {journal} {\bibinfo  {journal} {IEEE
  Trans. Signal Process}\ } (\bibinfo {year} {2008})}\BibitemShut {NoStop}%
\bibitem [{\citenamefont {Rilling}\ and\ \citenamefont
  {Flandrin}(2009)}]{Rilling2009}%
  \BibitemOpen
  \bibfield  {author} {\bibinfo {author} {\bibfnamefont {G.}~\bibnamefont
  {Rilling}}\ and\ \bibinfo {author} {\bibfnamefont {P.}~\bibnamefont
  {Flandrin}},\ }\href@noop {} {\bibfield  {journal} {\bibinfo  {journal} {Adv.
  Adapt. Data Anal.}\ }\textbf {\bibinfo {volume} {1}},\ \bibinfo {pages} {43}
  (\bibinfo {year} {2009})}\BibitemShut {NoStop}%
\bibitem [{\citenamefont {Schertzer}\ \emph {et~al.}(1997)\citenamefont
  {Schertzer}, \citenamefont {Lovejoy}, \citenamefont {Schmitt}, \citenamefont
  {Chigirinskaya},\ and\ \citenamefont {Marsan}}]{Schertzer1997}%
  \BibitemOpen
  \bibfield  {author} {\bibinfo {author} {\bibfnamefont {D.}~\bibnamefont
  {Schertzer}}, \bibinfo {author} {\bibfnamefont {S.}~\bibnamefont {Lovejoy}},
  \bibinfo {author} {\bibfnamefont {F.~G.}\ \bibnamefont {Schmitt}}, \bibinfo
  {author} {\bibfnamefont {Y.}~\bibnamefont {Chigirinskaya}}, \ and\ \bibinfo
  {author} {\bibfnamefont {D.}~\bibnamefont {Marsan}},\ }\href@noop {}
  {\bibfield  {journal} {\bibinfo  {journal} {Fractals}\ }\textbf {\bibinfo
  {volume} {5}},\ \bibinfo {pages} {427} (\bibinfo {year} {1997})}\BibitemShut
  {NoStop}%
\bibitem [{\citenamefont {Schmittbuhl}\ \emph {et~al.}(1995)\citenamefont
  {Schmittbuhl}, \citenamefont {Schmitt},\ and\ \citenamefont
  {Scholz}}]{Schmittbuhl1995JGR}%
  \BibitemOpen
  \bibfield  {author} {\bibinfo {author} {\bibfnamefont {J.}~\bibnamefont
  {Schmittbuhl}}, \bibinfo {author} {\bibfnamefont {F.~G.}\ \bibnamefont
  {Schmitt}}, \ and\ \bibinfo {author} {\bibfnamefont {C.}~\bibnamefont
  {Scholz}},\ }\href@noop {} {\bibfield  {journal} {\bibinfo  {journal} {J.
  geophys. Res}\ }\textbf {\bibinfo {volume} {100}},\ \bibinfo {pages} {5953}
  (\bibinfo {year} {1995})}\BibitemShut {NoStop}%
\bibitem [{\citenamefont {Schmitt}\ \emph {et~al.}(1995)\citenamefont
  {Schmitt}, \citenamefont {Lovejoy},\ and\ \citenamefont
  {Schertzer}}]{Schmitt1995GRL}%
  \BibitemOpen
  \bibfield  {author} {\bibinfo {author} {\bibfnamefont {F.~G.}\ \bibnamefont
  {Schmitt}}, \bibinfo {author} {\bibfnamefont {S.}~\bibnamefont {Lovejoy}}, \
  and\ \bibinfo {author} {\bibfnamefont {D.}~\bibnamefont {Schertzer}},\
  }\href@noop {} {\bibfield  {journal} {\bibinfo  {journal} {Geophys. Res.
  Lett.}\ }\textbf {\bibinfo {volume} {22}},\ \bibinfo {pages} {1689} (\bibinfo
  {year} {1995})}\BibitemShut {NoStop}%
\bibitem [{\citenamefont {O{\'s}wi\c{e}cimka}\ \emph
  {et~al.}(2006)\citenamefont {O{\'s}wi\c{e}cimka}, \citenamefont
  {Kwapie{\'n}},\ and\ \citenamefont {Dro{\.z}d{\.z}}}]{Oswicecimka2006PRE}%
  \BibitemOpen
  \bibfield  {author} {\bibinfo {author} {\bibfnamefont {P.}~\bibnamefont
  {O{\'s}wi\c{e}cimka}}, \bibinfo {author} {\bibfnamefont {J.}~\bibnamefont
  {Kwapie{\'n}}}, \ and\ \bibinfo {author} {\bibfnamefont {S.}~\bibnamefont
  {Dro{\.z}d{\.z}}},\ }\href@noop {} {\bibfield  {journal} {\bibinfo  {journal}
  {Phy. Rev. E}\ }\textbf {\bibinfo {volume} {74}},\ \bibinfo {pages} {16103}
  (\bibinfo {year} {2006})}\BibitemShut {NoStop}%
\bibitem [{\citenamefont {Heneghan}\ and\ \citenamefont
  {McDarby}(2000)}]{Heneghan2000PRE}%
  \BibitemOpen
  \bibfield  {author} {\bibinfo {author} {\bibfnamefont {C.}~\bibnamefont
  {Heneghan}}\ and\ \bibinfo {author} {\bibfnamefont {G.}~\bibnamefont
  {McDarby}},\ } {\bibfield  {journal}
  {\bibinfo  {journal} {Phys. Rev. E}\ }\textbf {\bibinfo {volume} {62}},\
  \bibinfo {pages} {6103} (\bibinfo {year} {2000})}\BibitemShut {NoStop}%
\bibitem [{\citenamefont {Chen}\ \emph {et~al.}(2002)\citenamefont {Chen},
  \citenamefont {Ivanov}, \citenamefont {Hu},\ and\ \citenamefont
  {Stanley}}]{Chen2002PRE}%
  \BibitemOpen
  \bibfield  {author} {\bibinfo {author} {\bibfnamefont {Z.}~\bibnamefont
  {Chen}}, \bibinfo {author} {\bibfnamefont {P.~C.}\ \bibnamefont {Ivanov}},
  \bibinfo {author} {\bibfnamefont {K.}~\bibnamefont {Hu}}, \ and\ \bibinfo
  {author} {\bibfnamefont {H.~E.}\ \bibnamefont {Stanley}},\ }\href@noop {}
  {\bibfield  {journal} {\bibinfo  {journal} {Phys. Rev. E}\ }\textbf {\bibinfo
  {volume} {65}},\ \bibinfo {pages} {041107} (\bibinfo {year}
  {2002})}\BibitemShut {NoStop}%
\bibitem [{\citenamefont {Koscielny-Bunde}\ \emph {et~al.}(2006)\citenamefont
  {Koscielny-Bunde}, \citenamefont {Kantelhardt}, \citenamefont {Braun},
  \citenamefont {Bunde},\ and\ \citenamefont {Havlin}}]{KoscielnyBunde2006}%
  \BibitemOpen
  \bibfield  {author} {\bibinfo {author} {\bibfnamefont {E.}~\bibnamefont
  {Koscielny-Bunde}}, \bibinfo {author} {\bibfnamefont {J.}~\bibnamefont
  {Kantelhardt}}, \bibinfo {author} {\bibfnamefont {P.}~\bibnamefont {Braun}},
  \bibinfo {author} {\bibfnamefont {A.}~\bibnamefont {Bunde}}, \ and\ \bibinfo
  {author} {\bibfnamefont {S.}~\bibnamefont {Havlin}},\ }\href@noop {}
  {\bibfield  {journal} {\bibinfo  {journal} {J. Hydrol.}\ }\textbf {\bibinfo
  {volume} {322}},\ \bibinfo {pages} {120} (\bibinfo {year}
  {2006})}\BibitemShut {NoStop}%
\bibitem [{\citenamefont {Sadegh~Movahed}\ \emph {et~al.}(2006)\citenamefont
  {Sadegh~Movahed}, \citenamefont {Jafari}, \citenamefont {Ghasemi},
  \citenamefont {Rahvar},\ and\ \citenamefont {Rahimi~Tabar}}]{Sadegh2006JSM}%
  \BibitemOpen
  \bibfield  {author} {\bibinfo {author} {\bibfnamefont {M.}~\bibnamefont
  {Sadegh~Movahed}}, \bibinfo {author} {\bibfnamefont {G.}~\bibnamefont
  {Jafari}}, \bibinfo {author} {\bibfnamefont {F.}~\bibnamefont {Ghasemi}},
  \bibinfo {author} {\bibfnamefont {S.}~\bibnamefont {Rahvar}}, \ and\ \bibinfo
  {author} {\bibfnamefont {M.}~\bibnamefont {Rahimi~Tabar}},\ }\href@noop {}
  {\bibfield  {journal} {\bibinfo  {journal} {J.  Stat. Mech.}\ ,\ \bibinfo {pages} {02003}} (\bibinfo {year}
  {2006})}\BibitemShut {NoStop}%
\bibitem [{\citenamefont {Bardet}\ and\ \citenamefont
  {Kammoun}(2008)}]{Bardet2008}%
  \BibitemOpen
  \bibfield  {author} {\bibinfo {author} {\bibfnamefont {J.}~\bibnamefont
  {Bardet}}\ and\ \bibinfo {author} {\bibfnamefont {I.}~\bibnamefont
  {Kammoun}},\ }\href@noop {} {\bibfield  {journal} {\bibinfo  {journal}
  {Information Theory, IEEE Transactions on}\ }\textbf {\bibinfo {volume}
  {54}},\ \bibinfo {pages} {2041} (\bibinfo {year} {2008})}\BibitemShut
  {NoStop}%
\bibitem [{\citenamefont {Zhang}\ \emph {et~al.}(2008)\citenamefont {Zhang},
  \citenamefont {Xu}, \citenamefont {Chen},\ and\ \citenamefont
  {Yu}}]{Zhang2008a}%
  \BibitemOpen
  \bibfield  {author} {\bibinfo {author} {\bibfnamefont {Q.}~\bibnamefont
  {Zhang}}, \bibinfo {author} {\bibfnamefont {C.}~\bibnamefont {Xu}}, \bibinfo
  {author} {\bibfnamefont {Y.}~\bibnamefont {Chen}}, \ and\ \bibinfo {author}
  {\bibfnamefont {Z.}~\bibnamefont {Yu}},\ }\href@noop {} {\bibfield  {journal}
  {\bibinfo  {journal} {Hydrol. Process.}\ }\textbf {\bibinfo {volume} {22}},\
  \bibinfo {pages} {4997} (\bibinfo {year} {2008})}\BibitemShut {NoStop}%
\bibitem [{\citenamefont {Mallat}\ and\ \citenamefont
  {Hwang}(1992)}]{Mallat1992singularity}%
  \BibitemOpen
  \bibfield  {author} {\bibinfo {author} {\bibfnamefont {S.}~\bibnamefont
  {Mallat}}\ and\ \bibinfo {author} {\bibfnamefont {W.}~\bibnamefont {Hwang}},\
  }\href@noop {} {\bibfield  {journal} {\bibinfo  {journal} {IEEE T. Inform.
  Theory.}\ }\textbf {\bibinfo {volume} {38}},\ \bibinfo {pages} {617}
  (\bibinfo {year} {1992})}\BibitemShut {NoStop}%
\bibitem [{\citenamefont {Mallat}(1999)}]{Mallat1999wavelet}%
  \BibitemOpen
  \bibfield  {author} {\bibinfo {author} {\bibfnamefont {S.}~\bibnamefont
  {Mallat}},\ }\href@noop {} { {\bibinfo {title} {{A wavelet tour of
  signal processing}}}}\ (\bibinfo  {publisher} {Academic Pr},\ \bibinfo {year}
  {1999})\BibitemShut {NoStop}%
\bibitem [{\citenamefont {Wendt}\ \emph {et~al.}(2007)\citenamefont {Wendt},
  \citenamefont {Abry},\ and\ \citenamefont {Jaffard}}]{Wendt2007}%
  \BibitemOpen
  \bibfield  {author} {\bibinfo {author} {\bibfnamefont {H.}~\bibnamefont
  {Wendt}}, \bibinfo {author} {\bibfnamefont {P.}~\bibnamefont {Abry}}, \ and\
  \bibinfo {author} {\bibfnamefont {S.}~\bibnamefont {Jaffard}},\ }\href@noop
  {} {\bibfield  {journal} {\bibinfo  {journal} {IEEE Signal Processing Mag.}\
  }\textbf {\bibinfo {volume} {24}},\ \bibinfo {pages} {38} (\bibinfo {year}
  {2007})}\BibitemShut {NoStop}%
\bibitem [{\citenamefont {Daubechies}(1992)}]{Daubechies1992}%
  \BibitemOpen
  \bibfield  {author} {\bibinfo {author} {\bibfnamefont {I.}~\bibnamefont
  {Daubechies}},\ }\href@noop {} { {\bibinfo {title} {{Ten lectures on
  wavelets}}}}\ (\bibinfo  {publisher} {Philadelphia: SIAM},\ \bibinfo {year}
  {1992})\BibitemShut {NoStop}%
\bibitem [{\citenamefont {Beran}(1994)}]{Beran1994}%
  \BibitemOpen
  \bibfield  {author} {\bibinfo {author} {\bibfnamefont {J.}~\bibnamefont
  {Beran}},\ }\href@noop {} { {\bibinfo {title} {{Statistics for
  long-memory processes}}}}\ (\bibinfo  {publisher} {CRC Press},\ \bibinfo
  {year} {1994})\BibitemShut {NoStop}%
\bibitem [{\citenamefont {Rogers}(1997)}]{Rogers1997}%
  \BibitemOpen
  \bibfield  {author} {\bibinfo {author} {\bibfnamefont {L.}~\bibnamefont
  {Rogers}},\ }\href@noop {} {\bibfield  {journal} {\bibinfo  {journal} {Math.
  Finance}\ }\textbf {\bibinfo {volume} {7}},\ \bibinfo {pages} {95} (\bibinfo
  {year} {1997})}\BibitemShut {NoStop}%
\bibitem [{\citenamefont {Doukhan}\ \emph {et~al.}(2003)\citenamefont
  {Doukhan}, \citenamefont {Taqqu},\ and\ \citenamefont
  {Oppenheim}}]{Doukhan2003}%
  \BibitemOpen
  \bibfield  {author} {\bibinfo {author} {\bibfnamefont {P.}~\bibnamefont
  {Doukhan}}, \bibinfo {author} {\bibfnamefont {M.}~\bibnamefont {Taqqu}}, \
  and\ \bibinfo {author} {\bibfnamefont {G.}~\bibnamefont {Oppenheim}},\
  }\href@noop {} {{\bibinfo {title} {{Theory and Applications of
  Long-Range Dependence}}}}\ (\bibinfo  {publisher} {Birkhauser},\ \bibinfo
  {year} {2003})\BibitemShut {NoStop}%
\bibitem [{\citenamefont {Gardiner}(2004)}]{Gardiner2004}%
  \BibitemOpen
  \bibfield  {author} {\bibinfo {author} {\bibfnamefont {C.~W.}\ \bibnamefont
  {Gardiner}},\ }\href@noop {} { {\bibinfo {title} {Handbook of Stochastic
  Methods}}}\ (\bibinfo  {publisher} {Springer, Berlin, third edition},\
  \bibinfo {year} {2004})\BibitemShut {NoStop}%
\bibitem [{\citenamefont {Wood}\ and\ \citenamefont {Chan}(1994)}]{Wood1994}%
  \BibitemOpen
  \bibfield  {author} {\bibinfo {author} {\bibfnamefont {A.}~\bibnamefont
  {Wood}}\ and\ \bibinfo {author} {\bibfnamefont {G.}~\bibnamefont {Chan}},\
  }\href@noop {} {\bibfield  {journal} {\bibinfo  {journal} {J. 
  Comput.   Graph. Stat.}\ }\textbf {\bibinfo {volume} {3}},\
  \bibinfo {pages} {409} (\bibinfo {year} {1994})}\BibitemShut {NoStop}%
\bibitem [{\citenamefont {Schmitt}(2003)}]{Schmitt2003}%
  \BibitemOpen
  \bibfield  {author} {\bibinfo {author} {\bibfnamefont {F.~G.}\ \bibnamefont
  {Schmitt}},\ }\href@noop {} {\bibfield  {journal} {\bibinfo  {journal} {Eur. Phys. J. B}\ }\textbf {\bibinfo {volume} {34}},\ \bibinfo
  {pages} {85} (\bibinfo {year} {2003})}\BibitemShut {NoStop}%
\bibitem [{\citenamefont {Bacry}\ \emph {et~al.}(2001)\citenamefont {Bacry},
  \citenamefont {Delour},\ and\ \citenamefont {Muzy}}]{Bacry2001}%
  \BibitemOpen
  \bibfield  {author} {\bibinfo {author} {\bibfnamefont {E.}~\bibnamefont
  {Bacry}}, \bibinfo {author} {\bibfnamefont {J.}~\bibnamefont {Delour}}, \
  and\ \bibinfo {author} {\bibfnamefont {J.F.}~\bibnamefont {Muzy}},\ }\href@noop
  {} {\bibfield  {journal} {\bibinfo  {journal} {Phys. Rev. E}\ }\textbf
  {\bibinfo {volume} {64}},\ \bibinfo {pages} {026103} (\bibinfo {year} {2001})}\BibitemShut {NoStop}%
\bibitem [{\citenamefont {Muzy}\ and\ \citenamefont {Bacry}(2002)}]{Muzy2002}%
  \BibitemOpen
  \bibfield  {author} {\bibinfo {author} {\bibfnamefont {J.F.}~\bibnamefont
  {Muzy}}\ and\ \bibinfo {author} {\bibfnamefont {E.}~\bibnamefont {Bacry}},\
  }\href@noop {} {\bibfield  {journal} {\bibinfo  {journal} {Phys. Rev. E}\
  }\textbf {\bibinfo {volume} {66}},\ \bibinfo {pages} {056121} (\bibinfo
  {year} {2002})}\BibitemShut {NoStop}%
\bibitem [{\citenamefont {Sreenivasan}(1991)}]{Sreenivasan1991}%
  \BibitemOpen
  \bibfield  {author} {\bibinfo {author} {\bibfnamefont {K.}~\bibnamefont
  {Sreenivasan}},\ }\href@noop {} {\bibfield  {journal} {\bibinfo  {journal}
  {Proc. R. Soc. Lond. A}\ }\textbf {\bibinfo {volume} {434}},\ \bibinfo
  {pages} {165} (\bibinfo {year} {1991})}\BibitemShut {NoStop}%
\bibitem [{\citenamefont {Shraiman}\ and\ \citenamefont
  {Siggia}(2000)}]{Shraiman2000}%
  \BibitemOpen
  \bibfield  {author} {\bibinfo {author} {\bibfnamefont {B.}~\bibnamefont
  {Shraiman}}\ and\ \bibinfo {author} {\bibfnamefont {E.}~\bibnamefont
  {Siggia}},\ }\href@noop {} {\bibfield  {journal} {\bibinfo  {journal}
  {Nature}\ }\textbf {\bibinfo {volume} {405}},\ \bibinfo {pages} {639}
  (\bibinfo {year} {2000})}\BibitemShut {NoStop}%
\bibitem [{\citenamefont {Warhaft}(2000)}]{Warhaft2000}%
  \BibitemOpen
  \bibfield  {author} {\bibinfo {author} {\bibfnamefont {Z.}~\bibnamefont
  {Warhaft}},\ }\href@noop {} {\bibfield  {journal} {\bibinfo  {journal} {Annu.
  Rev. Fluid Mech.}\ }\textbf {\bibinfo {volume} {32}},\ \bibinfo {pages} {203}
  (\bibinfo {year} {2000})}\BibitemShut {NoStop}%
\bibitem [{\citenamefont {Celani}\ \emph {et~al.}(2000)\citenamefont {Celani},
  \citenamefont {Lanotte}, \citenamefont {Mazzino},\ and\ \citenamefont
  {Vergassola}}]{Celani2000}%
  \BibitemOpen
  \bibfield  {author} {\bibinfo {author} {\bibfnamefont {A.}~\bibnamefont
  {Celani}}, \bibinfo {author} {\bibfnamefont {A.}~\bibnamefont {Lanotte}},
  \bibinfo {author} {\bibfnamefont {A.}~\bibnamefont {Mazzino}}, \ and\
  \bibinfo {author} {\bibfnamefont {M.}~\bibnamefont {Vergassola}},\
  }\href@noop {} {\bibfield  {journal} {\bibinfo  {journal} {Phys. Rev. Lett.}\
  }\textbf {\bibinfo {volume} {84}},\ \bibinfo {pages} {2385} (\bibinfo {year}
  {2000})}\BibitemShut {NoStop}%
\bibitem [{\citenamefont {Huang}\ \emph {et~al.}()\citenamefont {Huang},
  \citenamefont {Schmitt},\ and\ \citenamefont {Gagne}}]{Huang2011PRL}%
  \BibitemOpen
  \bibfield  {author} {\bibinfo {author} {\bibfnamefont {Y.}~\bibnamefont
  {Huang}}, \bibinfo {author} {\bibfnamefont {F.~G.}\ \bibnamefont {Schmitt}},
  \ and\ \bibinfo {author} {\bibfnamefont {Y.}~\bibnamefont {Gagne}},\
  }\href@noop {} {\bibinfo  {journal} {in preparation for Phys. Rev. Lett.}\
  }\BibitemShut {NoStop}%
\bibitem [{\citenamefont {Kolmogorov}(1941)}]{Kolmogorov1941}%
  \BibitemOpen
\bibfield  {journal} {  }\bibfield  {author} {\bibinfo {author} {\bibfnamefont
  {A.~N.}\ \bibnamefont {Kolmogorov}},\ }\href@noop {} {\bibfield  {journal}
  {\bibinfo  {journal} {Dokl. Akad. Nauk SSSR}\ }\textbf {\bibinfo {volume}
  {30}},\ \bibinfo {pages} {301} (\bibinfo {year} {1941})}\BibitemShut
  {NoStop}%
\bibitem [{\citenamefont {Obukhov}(1949)}]{Obukhov1949}%
  \BibitemOpen
  \bibfield  {author} {\bibinfo {author} {\bibfnamefont {A.}~\bibnamefont
  {Obukhov}},\ }\href@noop {} {\bibfield  {journal} {\bibinfo  {journal} {Izv.
  Acad. Nauk SSSR Ser. Geog. Geofiz}\ }\textbf {\bibinfo {volume} {13}},\
  \bibinfo {pages} {58} (\bibinfo {year} {1949})}\BibitemShut {NoStop}%
\bibitem [{\citenamefont {Corrsin}(1951)}]{Corrsin1951}%
  \BibitemOpen
  \bibfield  {author} {\bibinfo {author} {\bibfnamefont {S.}~\bibnamefont
  {Corrsin}},\ }\href@noop {} {\bibfield  {journal} {\bibinfo  {journal} {J.
  Appl. Phys.}\ }\textbf {\bibinfo {volume} {22}},\ \bibinfo {pages} {469}
  (\bibinfo {year} {1951})}\BibitemShut {NoStop}%
\bibitem [{\citenamefont {She}\ and\ \citenamefont
  {L{\'e}v{\^e}que}(1994)}]{She1994}%
  \BibitemOpen
  \bibfield  {author} {\bibinfo {author} {\bibfnamefont {Z.~S.}\ \bibnamefont
  {She}}\ and\ \bibinfo {author} {\bibfnamefont {E.}~\bibnamefont
  {L{\'e}v{\^e}que}},\ }\href@noop {} {\bibfield  {journal} {\bibinfo
  {journal} {Phys. Rev. Lett.}\ }\textbf {\bibinfo {volume} {72}},\ \bibinfo
  {pages} {336} (\bibinfo {year} {1994})}\BibitemShut {NoStop}%
\bibitem [{\citenamefont {van~de Water}\ and\ \citenamefont
  {Herwijer}(1999)}]{Water1999}%
  \BibitemOpen
  \bibfield  {author} {\bibinfo {author} {\bibfnamefont {W.}~\bibnamefont
  {van~de Water}}\ and\ \bibinfo {author} {\bibfnamefont {J.~A.}\ \bibnamefont
  {Herwijer}},\ }\href@noop {} {\bibfield  {journal} {\bibinfo  {journal} {J.
  Fluid Mech.}\ }\textbf {\bibinfo {volume} {387}},\ \bibinfo {pages} {3}
  (\bibinfo {year} {1999})}\BibitemShut {NoStop}%
\bibitem [{\citenamefont {Schmitt}(2006)}]{Schmitt2006}%
  \BibitemOpen
  \bibfield  {author} {\bibinfo {author} {\bibfnamefont {F.~G.}\ \bibnamefont
  {Schmitt}},\ }\href@noop {} {\bibfield  {journal} {\bibinfo  {journal}
  {Physica A}\ }\textbf {\bibinfo {volume} {368}},\ \bibinfo {pages} {377}
  (\bibinfo {year} {2006})}\BibitemShut {NoStop}%
\bibitem [{\citenamefont {Antonia}\ \emph {et~al.}(1984)\citenamefont
  {Antonia}, \citenamefont {Hopfinger}, \citenamefont {Gagne},\ and\
  \citenamefont {Anselmet}}]{Antonia1984}%
  \BibitemOpen
  \bibfield  {author} {\bibinfo {author} {\bibfnamefont {R.A.}~\bibnamefont
  {Antonia}}, \bibinfo {author} {\bibfnamefont {E.J.}~\bibnamefont {Hopfinger}},
  \bibinfo {author} {\bibfnamefont {Y.}~\bibnamefont {Gagne}}, \ and\ \bibinfo
  {author} {\bibfnamefont {F.}~\bibnamefont {Anselmet}},\ }\href@noop {}
  {\bibfield  {journal} {\bibinfo  {journal} {Phys. Rev. A}\ }\textbf {\bibinfo
  {volume} {30}},\ \bibinfo {pages} {2704} (\bibinfo {year}
  {1984})}\BibitemShut {NoStop}%
\bibitem [{\citenamefont {Ruiz-Chavarria}\ \emph {et~al.}(1996)\citenamefont
  {Ruiz-Chavarria}, \citenamefont {Baudet},\ and\ \citenamefont
  {Ciliberto}}]{Ruiz-Chavarria1996}%
  \BibitemOpen
  \bibfield  {author} {\bibinfo {author} {\bibfnamefont {G.}~\bibnamefont
  {Ruiz-Chavarria}}, \bibinfo {author} {\bibfnamefont {C.}~\bibnamefont
  {Baudet}}, \ and\ \bibinfo {author} {\bibfnamefont {S.}~\bibnamefont
  {Ciliberto}},\ }\href@noop {} {\bibfield  {journal} {\bibinfo  {journal}
  {Physica D}\ }\textbf {\bibinfo {volume} {99}},\ \bibinfo {pages} {369}
  (\bibinfo {year} {1996})}\BibitemShut {NoStop}%
\bibitem [{Note2()}]{Note2}%
  \BibitemOpen
  \bibinfo {note}
  {Http://perso.ens-lyon.fr/patrick.flandrin/emd.html}\BibitemShut {NoStop}%
\end{thebibliography}

\end{document}